\documentclass[aps,twocolumn,amsmath,amssymb,preprintnumbers]{revtex4}
\usepackage{amsmath} \usepackage{amsfonts} \usepackage{amssymb}
\usepackage{bbm}
\usepackage{epsfig}
\usepackage{graphics}
\usepackage{graphicx}
\newcommand{\be}{\begin{equation}}
\newcommand{\ee}{\end{equation}}
\newcommand{\ba}{\begin{eqnarray}}
\newcommand{\ea}{\end{eqnarray}}
\newcommand{\nn}{\nonumber}
\newcommand{\kr}{\rangle}
\newcommand{\kl}{\langle}

\newcommand{\cD}{{\cal D}}

\newcommand{\cN}{{\cal N}}
\newcommand{\cA}{{\cal A}}
\newcommand{\cK}{{\cal K}}
\newcommand{\cB}{{\cal B}}
\newcommand{\cP}{{\cal P}}
\newcommand{\cG}{{\cal G}}
\newcommand{\mn}{\mu\nu}
\newcommand{\bl}{\big(}
\newcommand{\br}{\big)}
\newcommand{\x}{(x)}
\newcommand{\xx}{(\bar x)}

\begin{document}

\title[ ]{Classical probabilities for Majorana and Weyl spinors}

\author{C. Wetterich}
\affiliation{Institut  f\"ur Theoretische Physik\\
Universit\"at Heidelberg\\
Philosophenweg 16, D-69120 Heidelberg}

\begin{abstract}
We construct a map between the quantum field theory of free Weyl or Majorana fermions and the probability distribution of a classical statistical ensemble for Ising spins or discrete bits. More precisely, a Grassmann functional integral based on a real Grassmann algebra specifies the time evolution of the real wave function $q_\tau(t)$ for the Ising states $\tau$. The time dependent probability distribution of a generalized Ising model obtains as $p_\tau(t)=q^2_\tau(t)$. The functional integral employs a lattice regularization for single Weyl or Majorana spinors. We further introduce the complex structure characteristic for quantum mechanics. Probability distributions of the Ising model which correspond to one or many propagating fermions are discussed explicitly. Expectation values of observables can be computed equivalently in the classical statistical Ising model or in the quantum field theory for fermions. 
\end{abstract}

\maketitle

\section{Introduction}
\label{Introduction}
It has recently been shown that quantum mechanics can be obtained from a classical statistical ensemble \cite{GR}. This requires a suitable law for the time evolution of the probability distribution $p_\tau(t)$ for the states $\tau$ of the classical statistical ensemble, and an appropriate selection of observables and their correlations describing sequences of measurements. An important criterion for the selection of the appropriate evolution law and observables is the compatibility with a ``coarse graining'' of the information to an isolated subsystem, for which only part of the information contained in $\{p_\tau(t)\}$ is available and used. 

The general setting how familiar ``no go theorems'' based on Bell's inequalities \cite{Bell}, \cite{BS} or the Kochen-Specker theorem \cite{KS} are circumvented has been discussed in \cite{GR} both on the abstract level and by simple concrete examples. For a practical use of the equivalence between quantum mechanics and a suitable classical statistical ensemble one needs an implementation for realistic systems, as a quantum particle in a potential or quantum field theories. We have presented a classical probability distribution in phase space which can describe a quantum particle in a potential, including all characteristic ``quantum features'' as interference in a double slit situation, tunneling and the discreteness of stationary states \cite{CWP}. For this purpose one has to postulate a specific ``unitary time evolution'' for the probability density in phase space which replaces the Liouville equation and is no longer compatible with the notion of particles following classical trajectories. While the choice of this evolution law can be motivated by the experimental success of quantum mechanics, one would also like to establish simple criteria for the fundamental time evolution law for probabilities in a classical statistical ensemble which do not rely anymore on the notion of individual trajectories. We believe that the answer to this problem is rooted on a more fundamental level, namely quantum field theory for many particle systems, from which the notion of a single particle in a potential emerges as a special case.

For the description of a quantum particle by a classical statistical ensemble in phase space the position and momentum observables are ``statistical observables''. The expectation values of statistical observables are computable from the probability distribution, despite the fact that statistical observables have no fixed value in the ``classical states'' (which correspond to points in phase space in this case). While statistical observables are familiar in classical statistical physics - an example is the entropy - they are usually not employed on a microscopic level. This raises the interesting question if the statistical observables on the level of one particle states can be obtained on a more fundamental level from classical observables which take a fixed value in every state $\tau$. It seems natural that this fundamental level corresponds to a quantum field theory. Indeed, we have demonstrated \cite{3A} that statistical observables and non-commutative products of observables can arise naturally from classical observables on a more fundamental level.

A natural goal for the embedding of quantum mechanics into a fundamental description by ``classical'' probabilities seems to be the explicit construction of a classical statistical ensemble whose probability distribution describes a quantum field theory. As a first step, we have constructed the ``classical'' probability distribution for a quantum field theory of free massless fermions in two dimensions \cite{CWF}. This  employs classical states $\tau$ which correspond to the states of Ising spins on a one-dimensional lattice or chain. For every point $x$ one introduces an occupation number $n(x)$ which can take the values one or zero. (Equivalently we could use the Ising spins $s(x)=2n(x)-1$ which have values $\pm 1$.) A classical state $\tau$ is a sequence of occupation numbers $\tau=\big\{n(x)\big\}$. The analogy to occupation numbers for fermions is very direct. Indeed, one has a map between states $\tau$ and basis elements $g_\tau$ of a real Grassmann algebra \cite{CWES} and one can associate the probability distribution $p_\tau(t)$ to a Grassmann functional integral \cite{CWF}. 

In this paper we take a second step and construct the probability distribution which describes a quantum field theory of massless Weyl or Majorana fermions in four dimensions. We will generalize this setting to Dirac spinors in an electromagnetic field in a forthcoming publication, but already the case of massless fermions has to circumvent a few obstacles and provides for important lessons. The basic setting is a generalization of ref. \cite{CWF}. The space points $\vec x$ are now on a cubic three dimensional lattice, and we employ four different occupation numbers $n_\gamma(\vec x), \gamma=1\dots 4$. The states of the classical ensemble are given by sequences of occupation numbers, $\tau=\{n_\gamma(\vec x)\}$. 

An important ingredient for the formulation of a fundamental evolution law for the probability distribution $\{p_\tau(t)\}$ is the ``classical'' wave function $q_\tau(t)$ \cite{CWP,CWF}. For every time $t$ this associates a real number $q_\tau$ to every state $\tau$. The probabilities $p_\tau$ are the squares of $q_\tau$ and therefore automatically positive, $p_\tau=q^2_\tau$. The normalization of the probability distribution requires
\be\label{AA}
\sum_\tau p_\tau=\sum_\tau q^2_\tau=1.
\ee
In turn, the wave function can be obtained from the probability distribution up to a sign $s_\tau=\pm 1$, $q_\tau=s_\tau\sqrt{p_\tau}$. For all quantities that can be computed from $\{p_\tau\}$ the choice of the signs $s_\tau$ corresponds to a choice of gauge without observable consequences. (Convenient gauge choices respect continuity and differentiability of $q_\tau(t)$, which excludes arbitrary ``jumps'' of $s_\tau$ and fixes $s_\tau$ to a large extent by the properties of $\{p_\tau\}$. Then $\{q_\tau\}$ is essentially computable from $\{p_\tau\}$, up to a few remaining ``overall signs'' \cite{CWP}.) The concept of the classical wave function permits the formulation of simple time evolution laws \cite{CWES} which preserve the normalization of the probabilities. It is sufficient that the time evolution is described by a rotation of the real vector $q_\tau$, 
\be\label{AB}
q_\tau(t')=\sum_\rho R_{\tau\rho}(t',t)q_\rho(t)~,~R^TR=1.
\ee
The simplest evolution laws are linear - the rotation matrix $R_{\tau\rho}(t',t)$ does not depend on the wave function $\{q_\tau(t)\}$. 

The classical wave function $\{q_\tau(t)\}$ is a property of the classical statistical ensemble and resembles in certain aspects to the Hilbert space formulation of classical mechanics by Koopman and von Neumann \cite{Kop}. In contrast to ref. \cite{Kop} it is, however, a real function. This avoids the introduction of additional degrees of freedom which would correspond to the phases of a complex wave function. On the other hand, the physics of phases is one of the most characteristic features of quantum physics. We therefore have to implement a complex structure which maps the real classical wave function $\{q_\tau(t)\}$ to an associated complex quantum wave function. For four dimensional fermions a natural complex structure can be associated to the equivalence between Majorana and Weyl spinors \cite{CWMS}, \cite{CWMSa}. Weyl spinors are directly formulated as complex entities, and the wave function for a single propagating Weyl fermion is indeed a complex function. On the other hand, a real representation of the Clifford algebra can be employed for Majorana fermions. The real wave function for a single particle can then be mapped to the complex wave function for Weyl spinors. 

We will extract the real wave function $\{q_\tau(t)\}$ and its time evolution from a Grassmann functional integral based on a real Grassmann algebra. The action will involve four ``real'' Grassmann fields $\psi_\gamma(t,x)$, but no separate conjugate fields $\hat\psi_\gamma(t,x)$ or $\bar\psi_\gamma(t,x)$. So far, the construction of the classical wave function from a Grassmann functional integral has been based on pairs of conjugate Grassmann variables $\psi$ and $\hat\psi$ \cite{CWF}. We have to generalize this construction for the case where no conjugate fields $\hat\psi$ are available. Furthermore, a well defined setting requires a regularization of the functional integral that we implement on a space-time lattice. Since we want to be able to describe single propagating Weyl or Majorana fermions we have to develop a discretization for single Weyl spinors without ``lattice doublers''. These constructions are all performed explicitly and we end with a real Grassmann functional integral which describes a single species of  massless Weyl or Majorana spinors in Minkowski space. We explicitly infer the wave functions for multi-particle and hole states. As usual in quantum mechanics, they depend on ``initial conditions'' set by the wave function $\{q_\tau(t_{in}\}$ a same initial time $t_{in}$. 

For a given action specifying the quantum field theory our construction allows the explicit computation of the quantum wave function $\{q_\tau(t)\}$ for an arbitrary initial wave function $\{q_\tau(t_{in})\}$. We may therefore ask what are the possibilities for static wave functions that are invariant under Poincar\'e transformations in the continuum limit. Any such wave function characterizes a possible ``vacuum state''. We find by explicit construction that the vacuum state is not unique for a quantum field theory of free massless fermions. For every possible vacuum state we can construct one-particle and one-hole excitations corresponding to single propagating fermions. Their wave function obeys a free Dirac equation. Also the construction of multi-fermion states by applying appropriate annihilation and creation operators to the vacuum state is the same for all possible vacuum states. 

The physical meaning of the degeneracy of vacuum states has not yet been eludicated. It is an interesting speculation that they cannot be distinguished by ``macroscopic observables''. In this case a ``coarse graining'' which averages over different vacuum states (and the associated excitations) would seem appropriate. Some light may be shed on this question by an investigation of massive Dirac spinors in an external electromagnetic field. Massless Dirac spinors can be understood as two species of massless Majorana spinors, but we do not know at the present stage the effect of the mass and the external potential on the degeneracy of the vacuum or more general ground states. 

For any given vacuum state we construct the one-particle wave function and discuss explicitly the corresponding probability distribution of the generalized Ising model. The time evolution of the wave function obeys the relativistic Dirac equation for a free massless Majorana particle or an equivalent Weyl-fermion. We specify observables for the position of the particle both as classical observables for the Ising-type model and as Grassmann operators. Expectation values of those observables can be computed equivalently from the classical statistical ensemble for the generalized Ising model, or from the Grassmann functional of the quantum field theory for fermions. We also discuss generalized one-particle states. For appropriate versions both momentum and position of a single particle can be formulated as classical observables. We introduce the non-commuting product of observables which underlies the measurement correlations of arbitrary functions of position and momentum.

This paper is organized as follows. We introduce in sect. \ref{Quantumfieldtheory} the action as an element of a real Grassmann algebra based on ``real'' Grassmann variables $\psi_\gamma(t,x)$. We employ a lattice discretization and establish the Lorentz-symmetry of the continuum limit. We also define a regularized Grassmann functional integral. Sect. \ref{Probabilitydistribution} discusses the map between this setting and the probability distribution of a generalized Ising model. The Grassmann functional integral permits to compute the probability distribution $\{p_\tau(t)\}$ for every time $t$. It obtains by ``integrating out'' the Grassmann variables for $t'>t$ and $t'<t$. 

In sect. \ref{Timeevolution} we turn to the time evolution of $\{p_\tau(t)\}$ and the associated ``classical wave function'' $\{q_\tau(t)\}$ which is induced by the action formulated in sect. \ref{Quantumfieldtheory}. We show that it is ``unitary'', corresponding to a rotation \eqref{AB} of the classical wave function with $R$ independent of $\{q\}$. We also investigate the notion of a conjugate wave function and use this to establish for the ``partition function'' the normalization $Z=1$. In the continuum limit the evolution equation becomes a type of Schr\"odinger equation for the real wave function $\{q_\tau(t)\}$ which leaves the norm of the wave function invariant. On the level of the classical probabilities $p_\tau(t)$ the evolution is non-linear. We also establish the evolution equation for the associated Grassmann wave function. In sect. \ref{Observables} we discuss the classical observables of the generalized Ising model in the usual setting of a classical statistical ensemble. They can be mapped to associated Grassmann operators, and the time evolution of expectation values can be described by a type of Heisenberg evolution equation. In particular, we discuss in this language the conserved quantities of our Ising type model. 

In sect. \ref{Particlestates} we turn to the interpretation of our system in terms of particles. The Ising model describes an arbitrary number of massless free propagating Majorana spinors or the associated holes. One particle states obey a Lorentz covariant Dirac equation. We translate this description of Majorana spinors to an equivalent description of Weyl spinors. In this language the one-particle wave function becomes complex. The general solutions for the one-particle or one-hole states are most easily discussed in this setting. The generalization to multi-fermion states is simple and leads to the usual totally antisymmetric wave functions for fermions. We discuss the associated symmetries, including continuous chiral symmetries and parity.

In sect. \ref{Complex structure} we discuss more systematically the complex structure which is associated to the equivalence between Majorana and Weyl spinors. We group the four ``real'' Grassmann variables $\psi_\gamma$ into two complex Grassmann variables $\zeta_\alpha$. This maps the real Grassmann algebra constructed from $\psi_\gamma$ to an associated complex Grassmann algebra based on $\zeta_\alpha$. The transformation between $\psi_\gamma$ and ($\zeta_\alpha,\zeta^*_\alpha)$ is a complex similarity transformation. Real Grassmann operators built from $\psi_\gamma$ and $\partial/\partial\psi_\gamma$ are mapped to complex Grassmann operators constructed from $\zeta_\alpha,\partial/\partial\zeta_\alpha$ etc.. Employing this complex structure, the action becomes the familiar complex action for free Weyl spinors. 

In sect. \ref{Particle-holeconjugation} we address the particle-hole conjugation. On the level of the classical Ising model this corresponds to an exchange of occupied and empty bits, or to a flip of sign of the Ising spins $s_\gamma(x)$. The particle-hole conjugation is realized by a map within the Grassmann algebra which is closely related to the exchange of conjugate elements. (It should not be associated to the usual charge conjugation in particle physics: Majorana spinors are invariant under charge conjugation or simply flip sign.) The particle-hole conjugation maps states with $n$ particles into states with $n$ holes. In particular, we have investigated possible vacuum states that are static, Lorentz-invariant and translation invariant and also symmetric under particle-hole conjugation. We show that with these conditions the vacuum state is not unique. In this section we also discuss position and momentum observables for various versions of one-particle states. Conclusions are presented in sect. \ref{Conclusionsanddiscussion}.

\section{Quantum field theory for fermions in four dimensions}
\label{Quantumfieldtheory}

In this section we formulate the quantum field theory for free Majorana spinors in four dimensions in terms of a Grassmann functional based on a real Grassmann algebra. The functional integral is regularized on a lattice. 

\bigskip\noindent
{\bf 1. Action}

\medskip
Let us consider the action
\be\label{F1}
S=\int_{t,x}\big\{\psi_\gamma\partial_t\psi_\gamma-\psi_\gamma
(T_k)_{\gamma\delta}
\partial_k\psi_\delta\big\}.
\ee
It involves four Grassmann functions $\psi_\gamma(t,x),\gamma=1\dots 4,x=(x_1,x_2,x_3)$. The integral extends over three dimensional space and time, with $\partial_t=\partial/\partial t$ and $\partial_k=\partial/\partial x_k$. The real symmetric matrices $T_k=(T_k)^T$ are given by
\ba\label{F2}
&&T_1=\left(\begin{array}{cccc}
0,&0,&1,&0\\0,&0,&0,&1\\1,&0,&0,&0\\0,&1,&0,&0
\end{array}
\right)
~,~
T_2=\left(\begin{array}{cccc}
0,&0,&0,&1\\0,&0,&-1,&0\\0,&-1,&0,&0\\1,&0,&0,&0
\end{array}
\right),\nn\\
&&T_3=\left(\begin{array}{cccc}
1,&0,&0,&0\\0,&1,&0,&0\\0,&0,&-1,&0\\0,&0,&0,&-1
\end{array}
\right),
\ea
and we deal with a real Grassmann algebra. Summation over repeated indices is implied. Within the Grassmann algebra the operation of transposition amounts to a total reordering of all Grassmann variables. The action \eqref{F1} is antisymmetric under this operation,
\be\label{F3}
S^T=-S.
\ee
If we define formally the ``Minkowski action'' $S_M=iS$, the latter is hermitean, $S_M=S^\dagger_M$, since $S^*_M=-S_M$. We will, however, not use the Minkowski action for our formulation.

\bigskip\noindent
{\bf   2. Lorentz symmetry}

\medskip
The action \eqref{F1} is invariant with respect to four dimensional Lorentz-transformations, as defined infinitesimally by $\psi_\gamma(t,x)\to\psi'_\gamma(t,x)+\delta\psi_\gamma(t,x)$, with $\epsilon_{\mu\nu}=-\epsilon_{\nu\mu},\mu=(0,k)$,
\be\label{F4}
\delta\psi_\gamma=-\frac12
\epsilon_{\mu\nu}\left(\Sigma^{\mu\nu}\right)_{\gamma\delta}\psi_\delta,
\ee
and $\psi'_\gamma(t,x)=\psi_\gamma(t,x)-(\xi^0\partial_t+\xi^k\partial_k)\psi_\gamma(t,x)$ accounting for the Lorentz transformed time and space coordinates. The matrices $\Sigma^{\mu\nu}$ are given by
\be\label{F5}
\Sigma^{0k}=-\frac12 T_k~,~\Sigma^{kl}=-\frac12\epsilon^{klm}\tilde I T_m,
\ee
with
\be\label{F6}
\tilde I=\left(\begin{array}{cccc}
0,&-1,&0,&0\\1,&0,&0,&0\\0,&0,&0,&-1\\0,&0,&1,&0
\end{array}\right)
=T_1T_2T_3.
\ee

The Lorentz invariance of the action $S$ is most easily established by employing the real matrices
\be\label{F7}
\gamma^0=\left(\begin{array}{cccc}
0,&0,&0,&1\\0,&0,&1,&0\\0,&-1,&0,&0\\-1,&0,&0,&0
\end{array}\right)~,~
\gamma^k=-\gamma^0 T_k,
\ee
such that
\be\label{F8}
S=-\int_{t,x}\bar\psi\gamma^\mu \partial_\mu \psi~,~\bar\psi=\psi^T\gamma^0,
\ee
where $\partial_0=\partial_t$ and $\bar\psi_\gamma=\psi_\delta(\gamma^0)_{\delta\gamma}$. (We consider $\psi$ here as a vector with components $\psi_\gamma$ and suppress the vector indices.) The real $4\times 4$ Dirac matrices $\gamma^\mu$ obey the Clifford algebra 
\be\label{F9}
\{\gamma^\mu,\gamma^\nu\}=2\eta^{\mu\nu},
\ee
with signature of the metric given by $\eta_{\mu\nu}=diag(-1,1,1,1)$. This can be easily verified by using the relations
\ba\label{F10}
\{T_k,T_l\}=2\delta_{kl}~,~
\{\gamma^0,T_k\}=0.
\ea
Furthermore, one finds
\be\label{F11}
(\gamma^0)^T=-\gamma^0~,~(\gamma^k)^T=\gamma^k,
\ee
and the relations
\ba\label{F12}
[T_k,T_l]&=&2\epsilon_{klm}\tilde I T_m~,~[T_k, \tilde I]=0~,~\tilde I^2=-1,\nn\\
\gamma^0\gamma^1\gamma^2\gamma^3&=&\tilde I~,~\{\gamma^0,\tilde I\}=0~,~
\{\gamma^k,\tilde I\}=0.
\ea
The Lorentz generators \eqref{F5} obtain from the Dirac matrices as
\be\label{F13}
\Sigma^{\mn}=-\frac14[\gamma^\mu,\gamma^\nu].
\ee
With
\be\label{F14}
\gamma^0(\Sigma^{\mn})^T\gamma^0=\Sigma^{\mn}
\ee
one finds that $\bar\psi$ transforms as
\be\label{F15}
\delta\bar\psi_\gamma=\frac12\epsilon_{\mn}
\bar\psi_\delta(\Sigma^{\mn})_{\delta\gamma},
\ee
such that $\bar\psi\psi=\bar\psi_\gamma\psi_\gamma$ is a Lorentz scalar. We recognize in eq. \eqref{F8} the standard Lorentz invariant action for free fermions in a Majorana representation of the Clifford algebra with real $\gamma^\mu$-matrices. 

\bigskip\noindent
{\bf  3. Discretization and functional integral}

\medskip
The action \eqref{F1} may be considered as the continuum limit of the regularized action
\be\label{N1}
S=\sum^{t_f-\epsilon}_{t=t_{in}}L(t),
\ee
with
\be\label{A17}
L(t)=\sum_x\psi_\gamma(t,x)B_\gamma(t+\epsilon,x).
\ee
We will place $x$ on points of a cubic lattice and consider a regularization where $\epsilon$ is small compared to the lattice distance, permitting a continuum limit $\epsilon\to 0$ for time even for discrete space points. Here $B_\gamma(t,x)$ is a linear combination of Grassmann variables $\psi(t,x)$, which we write as
\be\label{C.8}
B_\gamma(t,x)=\psi_\gamma(t,x)-\epsilon F_\gamma\big[\psi(t);x\big],
\ee
such that
\be\label{C.1}
L(t)=\sum_x\psi_\gamma(t,x)\Big\{\psi_\gamma(t+\epsilon,x)-\epsilon F_\gamma
\big[\psi(t+\epsilon);x\big]\Big\}.
\ee
The sum in eq. \eqref{N1} extends over discrete time points $t_n$, with $\int_t=\epsilon\sum_t=\epsilon\sum_n~,~t_{n+1}-t_n=\epsilon,n\in{\mathbbm Z}$, $t_{in}\leq t_n\leq t_f$. We employ
\be\label{C.2}
\partial_t\psi(t)=\frac1\epsilon\big[\psi(t+\epsilon)-\psi(t)\big],
\ee
such that the Grassmann property $\psi^2_\gamma(t,x)=0$ results in 
\be\label{C.3}
\psi_\gamma(t,x)\partial_t\psi_\gamma(t,x)=\frac1\epsilon ~ \psi_\gamma(t,x)\psi_\gamma(t+\epsilon,x).
\ee
The continuum limit is taken as $\epsilon\to 0$ for fixed $t_{in},t_f$. 

Similarly, we sum in eq. \eqref{C.1} over points $x$ of a cubic lattice with lattice distance $\Delta$ and $\int_x=\Delta^3\sum_x$. Correspondingly, $F_\gamma$ is defined as 
\ba\label{C.4}
F_\gamma\big[\psi(t+\epsilon);x\big]=
(T_k)_{\gamma\delta}\partial_k\psi_\delta(t+\epsilon,x)+0(\epsilon)
\ea
with use of the lattice derivative
\be\label{C.5}
\partial_k=\tilde\partial_k-\frac14\sum_{l,m}
\epsilon_{klm}T_lT_m\tilde I\gamma^0\tilde\delta_k.
\ee
Here we employ
\be\label{C.6}
\tilde\partial_k\psi_\gamma(x)=\frac{1}{2\Delta}\big[\psi_\gamma(x+\Delta_k)
-\psi_\gamma(x-\Delta_k)\big]
\ee
and
\be\label{C.7}
\tilde \delta_k\psi_\gamma(x)=\frac{1}{2\Delta}\big\{\psi_\gamma(x+\Delta_k)+\psi_\gamma
(x-\Delta_k)-2\psi_\gamma(x)\big\},
\ee
with $\Delta_k$ the unit lattice vector of length $\Delta$ in the $k$-direction, $\Delta^2_k=\Delta^2$. Details of the choice of the discretization and properties of the lattice derivative are discussed in appendix A. For our purpose it is important that the discretization does not introduce any fermion doubling, such that we can describe single Weyl or Majorana spinors. The term $\sim\tilde\delta_k$, which prevents the occurrence of doublers, vanishes in the continuum limit $\Delta\to 0$. 

In appendix A we also specify the omitted term $\sim 0(\epsilon)$ in eq. \eqref{C.4} which vanishes for $\epsilon\to 0$ or $\Delta\to 0$. This term is needed in order to guarantee that the Grassmann element $B_\gamma(t,x)$ has the same normalization as $\psi_\gamma(t,x)$. In other words, we can obtain $B_\gamma$ from $\psi_\gamma$ by a rotation
\ba\label{C.9}
&&B_\gamma(x)=\sum_y\bar R_{\gamma\delta}(x,y)\psi_\delta(y),\nn\\
&&\sum_y\bar R_{\eta\delta}(z,y)
\bar R_{\gamma\delta}(x,y)=\delta_{\eta\gamma}\delta(z,x).
\ea
We may take the limit $\epsilon\to 0$ first. At this stage the model is formulated on a discrete space lattice. The continuum limit in the space direction, $\Delta\to 0$, can be taken at the end. We observe that the space-continuum limit, leading to the action \eqref{F1}, also involves a rescaling of the Grassmann variables by a factor $\Delta^{(-3/2)}$. The Lorentz symmetry is realized in the continuum limit where both $\epsilon$ and $\Delta$ approach zero.

The functional integral is defined by the partition function
\be\label{N9}
Z=\int{\cal D}\psi\bar g_f\big[\psi(t_f)\big]e^{-S}
g_{in}\big[\psi(t_{in})\big],
\ee
with the functional measure
\be\label{N10}
\int{\cal D}\psi=\prod_{t,x}\int \big(d\psi_4(t,x)\dots d\psi_1(t,x)\big).
\ee
The boundary terms $g_{in}$ and $\bar g_f$ only depend on the Grassmann variables $\psi(t_{in})$ and $\psi(t_f)$, respectively. As we will see below, the boundary terms $\bar g_f$ and $g_{in}$ are related to each other, such that the functional integral \eqref{N9} is fully specified by the choice of $g_{in}$. 

\section{Probability distribution and wave function}
\label{Probabilitydistribution}
In this section we associate to the Grassmann functional integral \eqref{N9} a family of classical probability distributions $\{p_\tau(t)\}$. For every given time $t$ this probability distribution associates to each classical state $\tau$ a classical probability $p_\tau(t)$, which is positive and normalized
\be\label{N11}
p_\tau(t)\geq 0~,~\sum_\tau p_\tau(t)=1.
\ee
The positivity and normalization of the probabilities holds for an arbitrary choice of $g_{in}$, provided $\bar g_f$ is related to $g_{in}$ appropriately. For every given $g_{in}$ the probability distribution $\{p_\tau(t)\}$ is uniquely computable for all $t$, such that the functional integral \eqref{N9} also specifies the time evolution of the probability distribution. 

The states $\tau$ of the classical statistical ensemble are discrete if we use the formulation of the functional integral on a space lattice. If we assume a cubic lattice with $L^3$ lattice points we associate to every point $x$ four bits (or Ising spins) $n_\gamma(x)$ that can take the values $1$ or $0$. The states $\tau$ are then given by the sequences of bits, $\tau=\big\{n_\gamma(x)\big\}$, each $\tau$ corresponding to an ordered chain of $4L^3$ numbers $1$ or $0$. There are a total of $2^{4L^3}$ different states $\tau$.

\bigskip\noindent
{\bf 1.  Grassmann wave function}

\medskip
The construction of the probability distribution $\{p_\tau\}$ will rely on the isomorphism between states $\tau$ and the basis elements $g_\tau$ of a Grassmann algebra that can be constructed from the Grassmann variables $\psi_\gamma(x)$ for fixed $t$. Each basis element $g_\tau$ is a product of Grassmann variables
\be\label{N12}
g_\tau=\psi_{\gamma_1}(x_1)\psi_{\gamma_2}(x_2)\dots
\ee
which is ordered in some convenient way. To be specific, we define some linear ordering of the lattice points and place variables  with ``smaller $x$'' to the left, and for each $x$ place smaller $\gamma$ to the left. If a Grassmann element $g_\tau$ contains a given variable $\psi_\gamma(x)$ we may put the number $n_\gamma(x)$ in the sequence $\tau$ to $0$, while we take $n_\gamma(x)=1$ if the variable $\psi_\gamma(x)$ does not appear in the product \eqref{N12}. This specifies the map between the $2^{4L^3}$ independent basis elements $g_\tau$ of the Grassmann algebra and the states $\tau$. 

Our first step computes for the functional integral \eqref{N9} a Grassmann wave function $g(t)$, which is an element of the Grassmann algebra constructed from the Grassmann variables $\psi_\gamma(t,x)$ at given $t$. In order to define this Grassmann wave function we decompose the action \eqref{N1}
\ba\label{M1}
S&=&S_<+S_>,\nn\\
S_<&=&\sum_{t'<t}L(t')~,~S_>=\sum_{t'\geq t}L(t').
\ea
The wave function $g(t)$ obtains by integrating out all Grassmann variables for $t'<t$
\be\label{M2}
g(t)=\int {\cal D}\psi(t'<t)e^{-S_<}g_{in}.
\ee
We observe that $g(t)$ depends on the Grassmann variables $\psi(t)$. More precisely, it can be constructed from linear combinations of products of the Grassmann variables $\psi_\gamma(t,x)$ for fixed $t$. 

\bigskip\noindent
{\bf 2.   Classical probabilities and wave function}

\medskip
We next expand $g$ in terms of the basis elements $g_\tau$ of this algebra
\be\label{M3}
g(t)=\sum_\tau q_\tau(t)g_\tau\big[\psi(t)\big].
\ee
We would like to associate the real coefficients $q_\tau(t)$ with the classical wave function, such that the classical probabilities obtain as $p_\tau(t)=q^2_\tau(t)$. This requires for every $t$ the normalization $\sum_\tau q^2_\tau(t)=1$. We will show in the next section that this normalization condition is indeed obeyed, provided it holds for the initial wave function $g(t_{in})=g_{in}\big[\psi_\gamma(t_{in},x)\big]$. 

The conjugate wave function is defined as
\be\label{M4}
\tilde g(t)=\int {\cal D} \psi(t'>t)\bar g_f e^{-S_>}.
\ee
Again, this is an element of the Grassmann algebra constructed from $\psi(t)$. In terms of $g$ and $\tilde g$ the partition function reads
\be\label{M5}
Z=\int{\cal D}\psi(t)\tilde g(t)g(t),
\ee
where the Grassmann integration $\int {\cal D}\psi(t)$ extends only over the Grassmann variables $\psi_\gamma(t,x)$ for a given time $t$. The conjugate basis elements of the Grassmann algebra $\tilde g_\tau$ are defined \cite{CWF} by the relation
\be\label{M6}
\tilde g_\tau g_\tau=\prod_x\prod_\gamma\psi_\gamma(x)
\ee
(no sum over $\tau$) and the requirement that no variable $\psi_\gamma(x)$ appears both in $\tilde g_\tau$ and $g_\tau$. They obey
\be\label{M6A}
\int{\cal D}\psi(t)\tilde g_\tau\big[\psi(t)\big]g_\rho\big[\psi(t)\big]=\delta_{\tau\rho}.
\ee
Expanding
\be\label{M7}
\tilde g(t)=\sum_\tau\tilde q_\tau(t)\tilde g_\tau\big[\psi(t)\big]
\ee
yields
\be\label{M8}
Z=\sum_\tau\tilde q_\tau(t)q_\tau(t)
\ee
We will see in the next section that for a suitable choice of $\bar g_f$ the conjugate wave function obeys for all $t$ the relation $\tilde q_\tau(t)=q_\tau(t)$. Together with the normalization condition $\sum_\tau q^2_\tau=1$ this guarantees $Z=1$.

In consequence, we can express the classical probabilities $p_\tau(t)$ directly in terms of the Grassmann functional integral 
\be\label{M9}
p_\tau(t)=\int {\cal D}\psi\bar g_f{\cal P}_\tau(t)e^{-S}g_{in},
\ee
with ${\cal P}_\tau(t)$ a projection operator
\ba\label{M10}
{\cal P}_\tau(t)g_\rho\big[\psi(t)\big]
&=&g_\tau\big[\psi(t)\big]\delta_{\tau\rho},\nn\\
\int\cD\psi(t)\tilde g_\sigma\big[\psi(t)\big]\cP_\tau(t)g_\rho\big[\psi(t)\big]
&=&\delta_{\tau\sigma}\delta_{\tau\rho}
\ea
The formal expression of $\cP_\tau$ in terms of the Grassmann variables $\psi(t)$ and derivatives $\partial/\partial\psi(t)$ can be found in ref. \cite{3A}.

The interpretation of the Grassmann functional integral in terms of classical probabilities is based on the map $g(t)\to \big\{ p_\tau(t)\big\}$, which in turn is related to the map $\{ q_\tau\}\to \{p_\tau\}=\{ q^2_\tau\}$. The opposite direction of a map $\big\{p_\tau(t)\big\}\to g(t)$ is also possible, provided we choose an appropriate sign convention for $s_\tau=\pm 1$ in 
\be\label{G11}
q_\tau(t)=s_\tau(t)\sqrt{p_\tau(t)}.
\ee
This sign convention corresponds to a choice of gauge which does not matter if all expectation values and correlations can finally be expressed in terms of the probabilities $\big\{p_\tau(t)\}$. The map $g(t)\leftrightarrow\big\{q_\tau(t)\big\}$ is invertible. A given ``quantum state'' or equivalent classical statistical ensemble may be specified by the ``initial value'' at some time $t_0, g(t_0)$, or the associated wave function $\big\{q_\tau(t_0)\big\}$ or classical probability distribution $\{p_\tau(t_0)\big\}$. This is equivalent to the specification of $g_{in}=g(t_{in})$ in the functional integral. 

\section{Time evolution}
\label{Timeevolution}
In this section we compute the time evolution of the wave function $\big\{q_\tau(t)\big\}$ and the associated probability distribution $\{p_\tau(t)\big\}=\big\{q^2_\tau(t)\big\}$. This will lead to a type of Schr\"odinger equation for the real wave function $\big\{q_\tau(t)\big\}$, as well as an associated evolution equation for the Grassmann wave function $g(t)$. We will use the properties of this evolution equation in order to establish that the norm $\sum_\tau q^2_\tau(t)$ is conserved, such that  $\big\{q^2_\tau(t)\big\}$ can indeed be interpreted as a time dependent probability distribution. 

Due to the particular form of the action \eqref{F1}, which only involves one type of Grassmann variables (and no conjugate variables as in ref. \cite{CWF}) the discrete formulation of the functional integral is crucial. We will see that $g(t)$ jumps between two neighboring time points, while it is smooth between $g(t+2\epsilon)$ and $g(t)$. We will therefore distinguish between even and odd time points and use the definition \eqref{M3} of the wave function only for even times. For odd times we employ a different definition, which will guarantee the smoothness of the time evolution of $\big\{q_\tau(t)\big\}$ for both even and odd time points. In the continuum limit, $\epsilon\to 0$, the time evolution of the wave function is then described by a continuous rotation of the vector $\big\{q_\tau(t)\big\}$. In Appendix B we show that by a distinction between even and odd time points, with $\hat\psi_\gamma(t,x)=\psi_\gamma(t+\epsilon,x)$ for $t$ even, we can actually map our formalism to the formalism of ref. \cite{CWF} with conjugate spinors $\hat\psi$. 

The time evolution of the Grassmann wave function obeys
\be\label{P1}
g(t+\epsilon)=\int\cD\psi(t)e^{-L(t)}g(t).
\ee
This determines $\big\{q_\tau(t+\epsilon)\big\}$ in terms of $\big\{q_\tau(t)\big\}$. Thus the action \eqref{N1}, \eqref{C.1} specifies the dynamics how the probability distribution $\big\{p_\tau(t)\big\}$ evolves in time. The particular dynamics of a given model is determined by the term $\sim\epsilon F_\gamma$ in eq. \eqref{C.1}. This will play the role of the Hamiltonian in quantum mechanics. 

\bigskip\noindent
{\bf   1. Trivial time evolution}

\medskip
In order to understand the structure of the evolution law \eqref{P1} we first consider the case $F_\gamma=0$. For a vanishing Hamiltonian this should describe a trivial time evolution with a static probability distribution. For $F_\gamma=0$ the evolution equation \eqref{P1} reads
\be\label{P2}
g(t+\epsilon)=\int\cD\psi(t)\exp\big\{-\sum_x\sum_\gamma\psi_\gamma(t,x)
\psi_\gamma(t+\epsilon,x)\big\}g(t).
\ee
We may write $g(t)$ in a product form
\be\label{P3}
g(t)=\prod_x\prod_\gamma\big[a_\gamma(t,x)+b_\gamma(t,x)\psi_\gamma(t,x)\big],
\ee
with some fixed ordering convention of the factors assumed, e.g. smaller $\gamma$ to the left for given $x$, and some linear ordering of the lattice points, with ``lower'' points to the left. In the product form eq. \eqref{P2} yields
\ba\label{P4}
&&g(t+\epsilon)=\int\cD\psi(t)\prod_x\prod_\gamma\\
&&\Big\{\big[1-\psi_\gamma(t,x)\psi_\gamma(t+\epsilon,x)\big]
\big[a_\gamma(t,x)+b_\gamma(t,x)\psi_\gamma(t,x)\big]\Big\}\nn\\
&&\hspace{1.3cm}=\prod_x\prod_\gamma\Big\{b_\gamma(t,x)+\eta_\gamma a_\gamma(t,x)\psi_\gamma(t+\epsilon,x)\Big\}.\nn
\ea
Here we use the fact that each individual Grassmann integration $\int d\psi_\gamma(t,x)$ can be performed easily,
\be\label{P4A}
\int d\psi(1-\psi\varphi)(a+b\psi)=b-a\varphi,
\ee
and $\eta_\gamma=\pm 1$ results from the anticommuting properties of the Grassmann variables $\varphi$, with
\be\label{P5}
\eta_1=\eta_3=1~,~\eta_2=\eta_4=-1.
\ee

As a result, we can write
\be\label{P6}
g(t+\epsilon)=\sum_\tau q_\tau(t){\cal C}g_\tau \big[\psi_\gamma(t+\epsilon,x)\big],
\ee
where ${\cal C}g_\tau$ obtains from $g_\tau$ by the following replacements: (i) for every factor $\psi_\gamma(x)$ in $g_\tau$ one has a factor $1$ in ${\cal C}g_\tau$; (ii) for every pair $(x,\gamma)$ for which no $\psi_\gamma(x)$ is present in $g_\tau$ one inserts a factor $\eta_\gamma\psi_\gamma(x)$ in ${\cal C}g_\tau$. 
The ordering of the factors $\eta_\gamma\psi_\gamma(x)$ is the same as the ordering assumed in the product \eqref{P3}. This implies that we can indeed write the action of ${\cal C}$ on the product \eqref{P3} as
\be\label{P7}
{\cal C}g(t)=\prod_x\prod_\gamma\big(b_\gamma(t,x)+\eta_\gamma a_\gamma(t,x)\psi_\gamma (t,x)\big).
\ee

The conjugation operator ${\cal C}$ maps every basis element $g_\tau$ into its conjugate element $\tilde g_\tau$ up to a sign $\sigma_\tau=\pm 1$,
\be\label{P8}
{\cal C}g_\tau=\sigma_\tau\tilde g_\tau.
\ee
Applying ${\cal C}$ twice on the product \eqref{P3} multiplies each factor by $\eta_\gamma$. The factors $\eta_\gamma$ drop out due to the even number of minus signs, such that ${\cal C}$ is an involution, ${\cal C}^2=1$, or 
\be\label{P9}
{\cal C}^2 g_\tau=g_\tau.
\ee
This has the consequence that $g(t+2\epsilon)$ obtains from $g(t)$ by simply replacing the arguments $\psi_\gamma(t,x)$ by $\psi_\gamma(t+2\epsilon,x)$, or
\be\label{P10}
q_\tau(t+2\epsilon)=q_\tau(t).
\ee
The trivial static evolution of the wave function is realized between $t$ and $t+2\epsilon$. 

The jump between $g$ and ${\cal C}g$ for neighboring time points, $g(t+\epsilon)={\cal C}g(t)$, suggests the use of eq. \eqref{M3} for the definition of the wave function $\big\{q_\tau(t)\big\}$ only for ``even time points'', namely those that obey $t_n=t_{in}+2m\epsilon,m\in{\mathbbm N}$. For odd time points, $t_n=t_{in}+(2m-1)\epsilon$, we employ a different definition, namely
\be\label{P11}
g(t)=\sum_\tau q_\tau(t){\cal C}g_\tau\big[\psi(t)\big].
\ee
With this definition one finds that for $F_\gamma=0$ the wave function $\big\{q_\tau(t)\big\}$ is indeed time independent, $q_\tau(t)=q_\tau(t_{in})$.

Similarly, we use the definition \eqref{M7} for $\tilde q_\tau(t)$ only for even $t$, while for odd $t$ it is replaced by
\be\label{P12}
\tilde g(t)=\sum_\tau\tilde q_\tau(t)\tilde{\cal C}\tilde g_\tau\big[\psi(t)\big].
\ee
The map $\tilde {\cal C}$ acts similarly as ${\cal C}$, with $\eta_\gamma$ replaced by $\tilde \eta_\gamma=-\eta_\gamma,
\tilde\eta_1=\tilde\eta_3=-1,\tilde\eta_2=\tilde\eta_4=1$, and $\tilde{\cal C}^2=1$. One observes
\be\label{52A}
\tilde {\cal C}\tilde g_\tau=\eta_\tau \sigma_\tau g_\tau,
\ee
with $\eta_\tau=-1$ if $g_\tau$ contains an odd number of Grassmann variables, while $\eta_\tau=1$ for an even number. Relation \eqref{52A} follows from ${\cal C}^2=1$ or
\be\label{52B}
{\cal C}\tilde g_\tau=\sigma_\tau g_\tau,
\ee
together with the property that $\tilde{\cal C}$ obtains from ${\cal C}$ by an additional minus sign for every $\psi$-factor
\be\label{52C}
\tilde {\cal C}\tilde g_\tau=\eta_\tau{\cal C}\tilde g_\tau.
\ee
The relation \eqref{52A} implies
\be\label{52D}
\int{\cal D}\psi\tilde{\cal C}\tilde g_\tau{\cal C}g_\rho=\delta_{\tau\rho}
\ee
such that the identity \eqref{M8} holds both for $t$ even and $t$ odd. 

Employing the identity
\be\label{P13}
\tilde g(t-\epsilon)=\int\cD\psi(t)\tilde g(t)e^{-L(t-\epsilon)},
\ee
one finds for $F_\gamma=0$
\be\label{P14}
\tilde g(t-\epsilon)=\sum_\tau\tilde q_\tau(t)\tilde{\cal C}\tilde g_\tau\big[\psi_\gamma(t-\epsilon,x)\big],
\ee
and therefore a time independent $\tilde q_\tau(t)$. With $\tilde g(t_f)=\bar g_f$ we can choose $\bar g_f=\sum_\tau q_\tau(t_{in})\tilde g_\tau\big[\psi(t_f)\big]$ for $t_f$ even, or $\bar g_f=\sum_\tau q_\tau(t_{in})\tilde{\cal C}\tilde g_\tau\big[\psi(t_f)\big]$ for $t_f$ odd. Then the relation $\tilde q_\tau(t)=q_\tau(t)$ holds for all $t$. We conclude that for $F_\gamma=0$ the dynamics is trivial, with time independent $\tilde q_\tau(t)=q_\tau(t)$ and therefore a time independent probability distribution $\{p_\tau\}$. 

\bigskip\noindent
{\bf 2. Unitary time evolution}

\medskip
We next turn to the dynamics which is described by the Lorentz-invariant action \eqref{F1} or the corresponding regularized form \eqref{N1}, \eqref{C.1}, with $F_\gamma$ given by eq. \eqref{C.4}. In consequence, we have to replace in eq. \eqref{P2} $\psi_\gamma(t+\epsilon,x)$ by $B_\gamma(t+\epsilon,x)$ according to eq. \eqref{C.8}
\be\label{P14A}
L(t)=\sum_x\sum_\gamma\psi_\gamma(t,x)B_\gamma(t+\epsilon,x).
\ee
This implies (for even $t$) 
\be\label{P15}
g(t+\epsilon)=\sum_\tau q_\tau(t){\cal C}g_\tau\big[B_\gamma(t+\epsilon,x)\big].
\ee
Since $B_\gamma(x)$ is related to $\psi_\gamma(x)$ by a rotation \eqref{C.9}, it is straightforward to show that $g_\tau\big[B_\gamma(x)\big]$ is also connected to $g_\tau\big[\psi_\gamma(x)\big]$ by a rotation among the basis elements
\ba\label{P16}
g_\tau\big[B_\gamma(x)\big]&=&\sum_\rho
g_\rho\big[\psi_\gamma(x)\big]R_{\rho\tau},\nn\\
\sum_\rho R_{\tau\rho}R_{\sigma\rho}&=&\delta_{\tau\sigma}.
\ea
One infers
\ba\label{P17}
g(t+\epsilon)&=&\sum_{\tau,\rho}q_\tau(t){\cal C}g_\rho\big[\psi_\gamma(t+\epsilon)\big]R_{\rho\tau}\nn\\
&=&\sum_\tau q_\tau(t+\epsilon){\cal C}g_\tau\big[\psi_\gamma(t+\epsilon)\big]
\ea
with a rotated wave function 
\be\label{P18}
q_\tau(t+\epsilon)=\sum_\rho R_{\tau\rho} q_\rho(t).
\ee
An analogous argument leads to the same evolution \eqref{P18} for $t$ odd. 

Rotations preserve the length of the vector $\{q_\tau\}$ such that $\sum_\tau q^2_\tau(t)$ is independent of $t$. Choosing 
$g_{in}=\sum_\tau q_\tau(t_{in})g_\tau\big[\psi(t_{in})\big]$ with $\sum_\tau q^2_\tau(t_{in})=1$ one infers $\sum_\tau q^2_\tau(t)=1$ for all $t$. Therefore $\big\{p_\tau(t)\big\}=\big\{q^2_\tau(t)\big\}$ has indeed for all $t$ the properties of a probability distribution, namely positivity of all $p_\tau$ and the normalization $\sum_\tau p_\tau=1$. In analogy to quantum mechanics we call a time evolution which preserves the norm of $\big\{q_\tau(t)\big\}$ a ``unitary time evolution''. A unitary time evolution is crucial for the probability interpretation of the functional integral \eqref{N9}. 

\newpage\noindent
{\bf  3. Evolution of conjugate wave function}

\medskip
We next want to show the relation (for $t$ even) 
\be\label{P19}
\tilde g(t-\epsilon)=\sum_{\tau,\rho}\tilde q_\tau(t)\tilde{\cal C}R_{\tau\rho}\tilde g_\rho
\big[\psi_\gamma(t-\epsilon,x)\big].
\ee
The definition \eqref{P12} of the conjugate wave function $\tilde q$,
\be\label{P19A}
\tilde g(t-\epsilon)=\sum_\tau\tilde q_\tau(t-\epsilon)\tilde{\cal C}\tilde g_\tau\big[\psi_\gamma(t-\epsilon,x)\big],
\ee
then implies
\be\label{P20}
\tilde q_\tau(t-\epsilon)=\sum_\rho \tilde q_\rho(t)R_{\rho\tau},
\ee
such that
\be\label{P21}
\tilde q_\tau(t)=\sum_\rho R_{\tau\rho}\tilde q_\rho(t-\epsilon).
\ee
Comparing with eq. \eqref{P18} one infers that $q_\tau(t)$ and $\tilde q_\tau(t)$ obey the same evolution equation. (By an analogous argument eq. \eqref{P21} also holds for $t$ odd.) If $q$ and $\tilde q$ are equal for some particular time $t_0$, they will remain equal for all $t$. 

In order to show eq. \eqref{P19} we employ eq. \eqref{P13},
\ba\label{P22}
&&\tilde g(t-\epsilon)=\int\cD\psi(t)\tilde g(t)\exp
\big\{-\sum_x\psi_\gamma(t-\epsilon,x)B_\gamma(t,x)\big\}\nn\\
&&=\int\cD\psi(t)\tilde g(t)\exp 
\big\{-\sum_{x,y}\psi_\gamma(t-\epsilon,x)
\bar R_{\gamma\delta}(x,y)\psi_\delta(t,y)\big\}\nn\\
&&=\int{\cal D}\psi(t)\tilde g(t)\exp
\big\{-\sum_x\tilde B_\gamma(t-\epsilon,x)\psi_\gamma(t,x)\big\},
\ea
which replaces in eq. \eqref{P14} $\psi_\gamma(t-\epsilon,x)$ by $\tilde B_\gamma(t-\epsilon,x)$, 
\ba\label{P23}
\tilde B_\gamma(t,x)&=&\sum_y\psi_\delta(t,y)\bar R_{\delta\gamma}(y,x),\nn\\
&=&\sum_y(\bar R^{-1})_{\gamma\delta}(x,y)\psi_\delta(t,y),
\ea
leading to
\be\label{P24}
\tilde g(t-\epsilon)=\sum_\tau\tilde q(t)\tilde{\cal C}\tilde g_\tau
\big[\tilde B_\gamma(t-\epsilon,x)\big].
\ee
Eq. \eqref{P19} then follows from the relation
\be\label{P25}
\tilde g_\tau\big[\tilde B_\gamma(t,x)\big]=\sum_\rho R_{\tau\rho}\tilde g_\rho
\big[\psi_\gamma(t,x)\big].
\ee
Indeed, we can expand $\tilde g_\tau\big[\tilde B_\gamma(t,x)\big]$ in terms of the complete set of basis elements $\tilde g_\rho\big[\psi_\gamma(t,x)\big]$,
\be\label{P26}
\tilde g_\tau\big[\tilde B_\gamma(t,x)\big]=\sum_\rho A_{\tau\rho}
\tilde g_\rho\big[\psi_\gamma(t,x)\big],
\ee
with
\ba\label{P27}
A_{\tau\rho}&=&\int \cD\psi(t)\tilde g_\tau
\big[\tilde B_\gamma(t,x)\big]
g_\rho\big[\psi_\gamma(t,x)\big]\nn\\
&=&\int \cD\psi(t)\tilde g_\tau
\big[\psi_\gamma(t,x)\big]g_\rho\big[B_\gamma(t,x)\big]\\
&=&\sum_\sigma\int\cD\psi(t)\tilde g_\tau \big[\psi_\gamma(t,x)\big]
g_\sigma\big[\psi_\gamma(t,x)\big]R_{\sigma\rho}=R_{\tau\rho}.\nn
\ea
The second line uses the fact that $\bar R\tilde B=\psi$ and the invariance of the functional measure $\int \cD\psi(t)$ under rotations $\psi\to\bar R\psi$. This concludes the proof of eq. \eqref{P19}. 

\bigskip\noindent
{\bf  4. Partition function}

\medskip
If $\big\{\tilde q_\tau(t_0)\big\}$ equals $\big\{ q_\tau(t_0)\big\}$ for some time $t_0$ we can use $\tilde q_\tau(t)=q_\tau(t)$ for all $t$ and infer from eq. \eqref{M8}
\be\label{P28}
Z=\sum_\tau q^2_\tau(t).
\ee
As it should be, $Z$ remains invariant under rotations \eqref{P18} of the vector $\{q_\tau\}$ and is therefore independent of $t$. Inversely, we may actually use the fact that the relation \eqref{M8} holds for all $t$ in order to show that $\{\tilde q_\tau\}$ and $\{q_\tau\}$ must obey the same time evolution provided $q(t+\epsilon)$ obtains from $q(t)$ by a rotation. Assume 
\be\label{P28A}
q_\tau(t+\epsilon)=\sum_\rho A_{\tau\rho}q_\rho(t)~,~\tilde q_\tau(t+\epsilon)
=\sum_\rho\tilde A_{\tau\rho}\tilde q_\rho(t)
\ee
for some arbitrary regular matrices $A$ and $\tilde A$. The time independence of eq. \eqref{M8} then implies in a matrix notation
\be\label{P29}
Z=\tilde q^T(t+\epsilon)q(t+\epsilon)=
\tilde q^T(t)\tilde A^TAq(t)=
\tilde q^T(t) q(t).
\ee
This holds for arbitrary $q$ and $\tilde q$ (corresponding to arbitrary $g_{in},\bar g_f$) provided
\be\label{P30}
\tilde A^TA=1~,~\tilde A=(A^T)^{-1}.
\ee
If $A$ describes a rotation the matrices $\tilde A$ and $A$ coincide. The relation \eqref{P30} demonstrates that the time evolution of $q_\tau$ by a rotation \eqref{P18} is a sufficient and necessary condition for $\tilde q(t)=q(t)$ and therefore for the expression \eqref{M9} of the probability in terms of the Grassmann functional integral. If $A$ is not a rotation matrix, the relation $\tilde q(t)=q(t)$ cannot be maintained for arbitrary $t$ and arbitrary $q(t_0)$. 

These remarks allow for straightforward generalizations of our setting. Whenever we can write
\be\label{P31}
L(t)=\sum_x\psi_\gamma(t,x)B_\gamma(t+\epsilon,x),
\ee
with $B_\gamma(t+\epsilon,x)$ containing terms with an arbitrary odd number of Grassmann variables $\psi_\gamma(t+\epsilon,x)$, then eq. \eqref{P16} is sufficient to guarantee $\sum_\tau q^2_\tau=const.$ and $\tilde q(t)=q(t)$ provided $\tilde q(t_0)=q(t_0)$. 

\newpage  \noindent
{\bf   5. Boundary terms}

\medskip
The final point we have to settle in order to establish the normalization $Z=1$ and the expression \eqref{M9} concerns the equality of $\tilde q(t_0)$ and $q(t_0)$ for some arbitrary time $t_0$. This is achieved by a proper choice of the relation between the boundary terms $\bar g_f$ and $g_{in}$ in eq. \eqref{N9}. For this purpose we may imagine that we (formally) solve the evolution equation \eqref{P18} in order to compute $\big\{q_\tau(t_f)\big\}$ in terms of $\big\{q_\tau(t_{in}\big\}$, $\bar g_{in}=\sum_\tau q_\tau(t_{in}) g_\tau\big[\psi(t_{in})\big]$. Let us assume that $t_f$ is even such that 
\ba\label{P32}
g(t_f)&=&\sum_\tau q_\tau(t_f)g_\tau\big[\psi(t_f)\big], \\
\tilde g(t_f)&=&\sum_\tau\tilde q_\tau(t_f)\tilde g_\tau\big[\psi(t_f)\big]=\bar g_f.\label{P32a}
\ea
It is then sufficient to choose $\bar g_f$ such that $\tilde q_\tau(t_f)=q_\tau(t_f)$. Equivalently, we may specify the wave function $\big\{q_\tau(t_0)\big\}=\big\{\tilde q_\tau(t_0)\big\}$ at some arbitrary time $t_0$ and compute the corresponding $g_{in}$ and $\bar g_f$ by a solution of the evolution equation, using the fact that the rotation \eqref{P18} can be inverted in order to compute $q(t-\epsilon)$ form $q(t)$. 

\bigskip\noindent
{\bf   6. Continuous evolution equation}

\medskip
Finally, we cast the evolution law \eqref{P18} into the form of a differential time evolution equation by taking the limit $\epsilon\to 0$. We employ eq. \eqref{P15} consecutively for two time steps
\ba\label{P34}
\partial_tg(t)&=&\frac{1}{2\epsilon}\big[g(t+2\epsilon)-g(t)\big]\nn\\
&=&\frac{1}{2\epsilon}\sum_\tau q_\tau(t)
\Big\{g_\tau\big[\psi_\gamma(x)-2\epsilon
F_\gamma(x)\big]-g_\tau\big[\psi_\gamma(x)\big]\Big\}\nn\\
&=&\sum_{\tau,\rho}q_\tau(t)g_\rho\big[\psi_\gamma(x)\big]K_{\rho\tau}\nn\\
&=&\sum_\tau\partial_t q_\tau(t)g_\tau\big[\psi_\gamma(x)\big],
\ea
resulting in a Schr\"odinger type equation for the real wave function $\big\{q_\tau(t)\big\}$,
\be\label{P35}
\partial_t q_\tau(t)=\sum_\rho K_{\tau\rho} q_\rho(t).
\ee
(In eq. \eqref{P34} we use a fixed basis, corresponding to the basis elements $g_\tau$ constructed from $\psi(t)$ for $g(t)$, and from $\psi(t+2\epsilon)$ for $g(t+2\epsilon)$.) Since the evolution describes a rotation, the matrix $K$ is antisymmetric
\be\label{P36}
K_{\rho\tau}=-K_{\tau\rho}.
\ee
We identify this evolution equation with the Schr\"odinger equation for a quantum wave function for the special case of a real wave function and purely imaginary and hermitean Hamiltonian $H=i\hbar K$. 

The time evolution \eqref{P35} translates directly to the probabilities (no summation over $\tau$ here)
\be\label{G14}
\partial_t p_\tau=2\sum_\rho K_{\tau\rho}s_\tau s_\rho\sqrt{p_\tau p_\rho}.
\ee
Once the signs $s_\tau(t_0)$ are fixed by some appropriate convention at a given time $t_0$, the signs $s_\tau(t)$ are computable in terms of the probabilities. This follows since for all $t$ the wave function $q_\tau(t)$ is uniquely fixed by $q_\tau(t_0)$ or $p_\tau(t_0)$, and $p_\tau(t)$ is uniquely determined by $q_\tau(t)$. In principle, it is therefore possible to formulate the time evolution law for the probabilities uniquely in terms of the probabilities.  In particular, the positive roots $\sqrt{p_\tau}$ obey 
\be\label{89A}
\partial_t\sqrt{p_\tau}=\sum_\rho s_\tau s_\rho K_{\tau\rho}\sqrt{p_\rho}.
\ee
This is a standard linear equation for $\sqrt{p_\tau}$, except for the additional sign information stored in $\{s_\tau (t)\}$. The role of the sign is to prevent $\sqrt{p_\tau}$ to become negative during the evolution and to maintain the normalization condition $\sum_\tau p_\tau=1$. Whenever for a fixed sign $s_\tau$ the evolution \eqref{89A} would lead to $\sqrt{p_\tau}$ crossing zero, the sign $s_\tau$ flips such that $\sqrt{p_\tau}$ increases in the following times instead of decreasing to negative values. In principle, this provides for an algorithm of keeping track of the appropriate signs $\{s_\tau(t)\}$. Whenever $p_\tau(t_0)=0$ for some time $t_0$, only one choice of $\sigma_\tau(t_0+\epsilon)$ is compatible with $p_\tau(t_0+\epsilon)\geq 0$ and $\sum_\tau p_\tau(t_0+\epsilon)=1$. This condition decides if $\sigma_\tau$ jumps at $t_0$ or not. 

The elegant way of keeping track of the sign information is, of course, the use of the wave function. The linear evolution law \eqref{P35} is very convenient for the description of the classical ensembles which obey an evolution law of the type \eqref{89A}. More generally, the general form of the evolution law \eqref{P35} allows for the construction of simple time evolution laws for classical probability distributions. For a general evolution law the basic property to be respected is the condition of unit norm of the probability distribution. This can be quite cumbersome for a general evolution equation for $\{p_\tau\}$, but it is extremely simple on the level of $\{q_\tau\}$ where only the length of a real vector has to be preserved. 

\bigskip\noindent
{\bf   7. Grassmann evolution equation}

\medskip
The matrix $K$ can be extracted from the Grassmann evolution equation
\be\label{P37}
\partial_t g={\cal K} g,
\ee
with Grassmann operator ${\cal K}$ given by
\be\label{P38}
{\cal K}=\sum_x\frac{\partial}{\partial\psi_\gamma(x)}(T_k)_{\gamma\delta}
\partial_k\psi_\delta(x).
\ee
In order to proof the relation \eqref{P37}, \eqref{P38} we consider the Grassmann operator
\ba\label{P39}
{\cal W}_\gamma(x)&=&\frac{\partial}{\partial\psi_\gamma(x)}\psi_\gamma(x)+
\big[\psi_\gamma(x)-2\epsilon F_\gamma(x)\big]
\frac{\partial}{\partial\psi_\gamma(x)}\nn\\
&=&1-2\epsilon F_\gamma(x)\frac{\partial}{\partial\psi_\gamma(x)}.
\ea
Its action on an arbitrary Grassmann element $g$ results in the replacement of the particular Grassmann variable $\psi_\gamma(x)$ by $\psi_\gamma(x)-2\epsilon F_\gamma(x)$. In terms of these operators we can write
\be\label{P40}
g(t+2\epsilon)=\big(\prod_{x,\gamma}{\cal W}_\gamma(x)\big)g(t)={\cal U}
(t+2\epsilon,t)g(t),
\ee
with ${\cal U}$ the Grassmann evolution operator \cite{CWF}, \cite{3A}. (In eq. \eqref{P40} we only have kept contributions up to linear order in $\epsilon$.) We can now verify the relations
\ba\label{P41}
{\cal K}&=&\sum_{x,\gamma}{\cal K}_\gamma(x),\nn\\
{\cal K}_\gamma(x)&=&-(T_k)_{\gamma\delta}\partial_k\psi_\delta(x)
\frac{\partial}{\partial\psi_\gamma(x)},\nn\\
{\cal W}_\gamma(x)&=&\exp\big\{2\epsilon{\cal K}_\gamma(x)\big\},
\ea
such that eq. \eqref{P40} results in 
\be\label{P42}
g(t+2\epsilon)=\exp(2\epsilon{\cal K})g(t)=g(t)+2\epsilon{\cal K} g(t),
\ee
thus establishing eq. \eqref{P37}. (Again, at several steps we have neglected corrections $\sim\epsilon^2$ that vanish in the continuum limit.) 

From the expansion of ${\cal K} g_\tau$ in the basis elements $g_\rho$,
\be\label{P43}
{\cal K} g_\tau=\sum_\rho g_\rho K_{\rho\tau},
\ee
we infer the matrix element
\be\label{P44}
K_{\rho\tau}=\int\cD\psi\tilde g_\rho{\cal K} g_\tau.
\ee
Taking finally the space-continuum limit by rescaling $\psi_\gamma(x)$ and $\partial/\partial\psi_\gamma(x)$ such that 
\be\label{P45}
\left\{\frac{\partial}{\partial\psi_\gamma(x)}~,~
\psi_\delta(y)\right\}=
\delta_{\gamma\delta}\delta^3(x-y),
\ee
we arrive at the continuum form of the Grassmann evolution equation
\be\label{P46}
\partial_tg={\cal K} g~,~{\cal K}=\int_x
\frac{\partial}{\partial\psi_\gamma(x)}(T_k)_{\gamma\delta}
\partial_k\psi_\delta(x).
\ee
This evolution equation will be the basis for the interpretation of the time dependent wave function $\big\{q_\tau(t)\big\}$ and probability distribution $\big\{p_\tau(t)\big\}$ in terms of propagating fermionic particles in section \ref{Particlestates}.

\section{Observables}
\label{Observables}

Classical observables $A$ take a fixed value $A_\tau$ for every classical state $\tau$. In classical statistics the possible outcomes of measurements of $A$ correspond to the spectrum of possible values $A_\tau$. The expectation value of $A$ obeys 
\be\label{G15}
\kl A\kr=\sum_\tau p_\tau A_\tau.
\ee
Our description of the system will be based on these classical statistical rules. For example, we may consider the observable measuring the occupation number $N_\gamma(x)$ of the bit $\gamma$ located at $x$. The spectrum  of possible outcomes of measurements consists of values $1$ or $0$, depending if a given state $\tau=[n_\gamma(x)]$ has this particular bit occupied or empty. 

\bigskip\noindent
{\bf   1. Grassmann operators}

\medskip
For the Grassmann basis element $g_\tau$ associated to $\tau$ one finds $N_\gamma(x)=0$ if $g_\tau$ contains a factor $\psi_\gamma(x)$, and $N_\gamma(x)=1$ otherwise. We can associate to this observable a Grassmann operator ${\cal N}_\gamma(x)$ obeying (no summation here) 
\be\label{G16}
\cN_\gamma(x)g_\tau=\big(N_\gamma(x)\big)_\tau g_\tau~,~\cN_\gamma(x)=
\frac{\partial}{\partial\psi_\gamma(x)}
\psi_\gamma(x).
\ee
Two occupation number operators ${\cal N}_{\gamma_1}(x_1)$ and ${\cal N}_{\gamma_2}(x_2)$ commute if $\gamma_1\neq \gamma_2$ or $x_1\neq x_2$. 

In general, we may associate to each classical observable $A$ a diagonal quantum operator $\hat A$ acting on the wave function, defined by
\be\label{G16a}
(\hat A q)_\tau=A_\tau q_\tau.
\ee
This yields the quantum rule for expectation values
\be\label{G17}
\kl A\kr=\kl q\hat A q\kr=\sum_{\tau,\rho}q_\tau\hat A_{\tau\rho}q_\rho,
\ee
with $\hat A$ a diagonal operator $\hat A_{\tau\rho}=A_\tau\delta_{\tau\rho}$. In the Grassmann formulation one uses the associated Grassmann operator ${\cal A}$ obeying
\be\label{G18}
\cA g_\tau=A_\tau g_\tau
\ee
such that 
\be\label{G19}
\kl A\kr=\int \cD\psi\tilde g\cA g.
\ee
Here $\tilde g$ is conjugate to $g$, i.e. for $g=\sum_\tau q_\tau g_\tau$ one has $\tilde g=\sum_\tau q_\tau\tilde g_\tau$. 

\bigskip\noindent
{\bf   2. Time evolution of expectation values}

\medskip
In classical statistics the time evolution of the expectation value is induced by the time evolution of the probability distribution
\be\label{95A} 
\kl A(t)\kr=\sum_\tau p_\tau(t)A_\tau.
\ee
This corresponds to the Schr\"odinger picture in quantum mechanics
\be\label{95B}
\kl A(t)\kr=\kl q(t)\hat A q(t)\kr,
\ee
or the corresponding expression in terms of the Grassmann algebra
\be\label{95C}
\kl A(t)\kr=\int \cD\psi\tilde g(t){\cal A}g(t).
\ee

Using $\partial_t q=Kq$ \eqref{P35}  we infer for the time evolution of the expectation value the standard relation
\be\label{95D}
\partial_t\kl A\kr=\kl q[\hat A,K]q\kr.
\ee
Similarly, we use $\partial_t g={\cal K}g$ \eqref{P37} and 
\be\label{95E}
\partial_t\tilde g=-{\cal K}^T\tilde g,
\ee
with ${\cal K}^T$ obeying for arbitrary Grassmann elements $f$ and $g$ the relation
\be\label{95F}
\int\cD\psi{\cal K}^T\tilde g f=\int\cD\psi\tilde g{\cal K}f.
\ee
One infers
\be\label{95G}
\partial_t\kl A\kr=\int\cD\psi\tilde g[{\cal A},{\cal K}]g.
\ee

\bigskip\noindent
{\bf   3. Conserved quantities}

\medskip
Conserved quantities correspond to Grassmann operators $\cA$ that commute with $\cK$. Let us consider operators of the type
\be\label{H1}
\cB_{\epsilon\eta}(y)=\frac{\partial}{\partial\psi_\epsilon(y)}~B_{\epsilon\eta}(y)
\psi_\eta(y),
\ee
with $B_{\epsilon\eta}$ depending on $y$ and derivatives with respect to $y$. They obey the commutator relation
\ba\label{H2}
&&\big[\cB_{\epsilon\eta}(y),\cK\big]=\partial_k
\left\{\frac{\partial}{\partial\psi_\gamma}(T_k)_{\gamma\epsilon}B_{\epsilon\eta}
\psi_\eta\right\}\\
&&\quad +\frac{\partial}{\partial\psi_\gamma}\big\{B_{\gamma\epsilon}
(T_k)_{\epsilon\eta}\partial_k\psi_\eta-(T_k)_{\gamma\epsilon}\partial_k(B_{\epsilon\eta}\psi_\eta)\big\},\nn
\ea
where all quantities on the r.h.s. depend on $y_k$ and $\partial_k=\partial/\partial y_{k}$. We can use eq. \eqref{H2} in order to find the conserved quantities
\be\label{H3a}
\cB_1=\int_y\frac{\partial}{\partial\psi_\gamma}b_1(\partial_k)\psi_\gamma~,~
\cB_2=\int_y\frac{\partial}{\partial\psi_\gamma}\tilde I_{\gamma\delta}b_2(\partial_k)
\psi_\delta.
\ee
Here we use the property that $\tilde I$ commutes with $T_k$ and $b_{1,2}$ contain only derivatives with respect to $y$. Hermitean (symmetric) operators are found if $b_1$ only involves even numbers of derivatives and $b_2$ only odd numbers. We observe the presence of off-diagonal conserved quantities as
\be\label{H4}
\cP_k=-\int_y\frac{\partial}{\partial\psi}\tilde I\partial_k\psi,
\ee
which are not classical statistical observables with fixed values in every state $\tau$, but can nevertheless be useful for  the understanding of the system.

\section{Particle states}
\label{Particlestates}

Our system admits a conserved particle number, corresponding to the Grassmann operator $\cN$,
\be\label{H5}
\cN=\int_y\frac{\partial}{\partial\psi_\gamma(y)}\psi_\gamma(y)~,~[\cN,\cK]=0.
\ee
The particle number is Lorentz invariant. We can decompose an arbitrary Grassmann element into eigenstates of $\cN$
\be\label{H6}
g=\sum_mA_mg_m~,~\cN g_m=mg_m.
\ee
The time evolution does not mix sectors with different particle number $m$, such that the coefficients $A_m$ are time independent
\be\label{H7}
\partial_t g=\sum_mA_m\partial_tg_m~,~\partial_tg_m=\cK g_m.
\ee
We can restrict our discussion to eigenstates of $\cN$.

\bigskip\noindent
{\bf   1. Vacuum}

\medskip
Let us consider some static vacuum state $g_0$ with a fixed particle number $m_0$,
\be\label{H8}
\cK g_0=0~,~\cN g_0=m_0g_0~,~\int \cD\psi \tilde g_0 g_0=1.
\ee
An example for a possible vacuum state is the totally empty state $g_0=|0\kr$, with
\ba\label{47Aa}
|0\kr=\prod_\alpha\psi_\alpha&=&\prod_x\prod_\gamma\psi_\gamma(x)
=\prod_x(\psi_1\psi_2\psi_3\psi_4),\nn\\
\cN|0\kr&=&0.
\ea
It obeys
\be\label{104Aa}
\int\cD\psi|0\kr=1~,~|\tilde 0\kr=1.
\ee
Another example is the totally occupied state $g_0=1$, with $m_0=B$. We shift the particle number by an additive ``renormalization'' $n=m-m_0$, such that the vacuum corresponds to $n=0$, and $g=A_n g_n$. An eigenstate of ${\cal N}$ with eigenvalue $m=m_0+n$ is called a $n$-particle state if $n$ is positive, and a $n$-hole state for negative $n$. 

We recall that in a formulation with discrete time steps $g(t)$ refers to even $t$. For odd $t$ and continuous time evolution the Grassmann wave function is given by $g(t_{odd})={\cal C}g(t)$. The conjugation ${\cal C}$ interchanges the role of particles and holes. Thus for $g(t)=|0\kr$ one has $g(t_{odd})=|1\kr=1$, such that the vacuum Grassmann wave function flips between fully empty and fully occupied for $t$ even and odd. Similarly, the excitations flip between particles and holes. Possible vacuum states with the same Grassmann wave function for $t$ even and odd, as $g_0=(|0\kr+|1\kr)\sqrt{2}$, will be discussed in sect. \ref{Particle-holeconjugation}. 

\bigskip\noindent
{\bf   2. One-particle and one-hole states}

\medskip
We next define creation and and annihilation operators $a^\dagger_\gamma(x),~a_\gamma(x)$ as
\be\label{47A}
a^\dagger_\gamma(x)g=\frac{\partial}{\partial\psi_\gamma(x)}g~,~a_\gamma(x)g=
\psi_\gamma(x)g.
\ee
They obey the standard (anti-)commutation relations
\ba\label{H6a}
\big \{a^\dagger_\gamma(x),~a_\epsilon(y)\big\}&=&\delta_{\gamma\epsilon}\delta(x-y)~,~
\cN=\int_xa^\dagger_\gamma(x)a_\gamma(x),\nn\\
~[a^\dagger_\gamma(x),\cN]&=&-a^\dagger_\gamma(x)~,~[a_\gamma(x),\cN]=a_\gamma(x).
\ea
Acting with the creation operator on the vacuum produces one-particle states
\ba\label{H7a}
&&g_1(t)=\int_xq_\gamma(t,x)a^\dagger_\gamma(x)g_0=\cG_1g_0,\nn\\
&&(\cN-m_0)g_1=g_1,
\ea
while a one-hole state with $n=-1$ obtains by employing the annihilation operator
\ba\label{H8a}
g_{-1}(t)&=&\int_x\hat q_\gamma(t,x)a_\gamma(x)g_0\nn\\
&=&-\int_x\bar q_\gamma(t,x)(\gamma^0)_{\gamma\delta}a_\delta(x)g_0=
\cG_{-1}g_0,\nn\\
&&~~(\cN-m_0)g_{-1}=-g_{-1},
\ea
with
\be\label{127A}
\bar q_\gamma=\hat q_\delta(\gamma^0)_{\delta\gamma}.
\ee

If we transform the one-particle wave function $q_\gamma(t,x)$ infinitesimally according to 
\ba\label{H9}
\delta q_\gamma=-\frac12\epsilon_{\mn}\left (\Sigma^{\mn}\right)_{\gamma\delta}q_\delta
\ea
the operator $\cG_1$ is Lorentz invariant. (We omit here the part resulting from the change of coordinates.) Similarly, $\hat q_\gamma$ transforms as $q_\gamma$ and the corresponding infinitesimal transformation
\be\label{H10}
\delta\bar q_\gamma=\frac12\epsilon_{\mn}
\bar q_\delta\left(\Sigma^{\mn}\right)_{\delta\gamma}
\ee
results in an invariant ${\cal G}_{-1}$. If the vacuum $g_0$ is Lorentz invariant, the Lorentz transformations \eqref{H9}, \eqref{H10} of the one-particle or one-hole wave functions will obey the same evolution equations as the original wave functions.

The time evolution of the one particle wave function $q_\gamma$ is given by
\be\label{H11} 
\partial_t g_1=\cK g_1=\int_x(\partial_tq\frac{\partial}{\partial\psi})g_0=\int_x 
q\left[\cK,\frac{\partial}{\partial\psi}\right]g_0.
\ee
and similar for the hole. With
\be\label{54A}
\left[\cK,\frac{\partial}{\partial\psi_\gamma(x)}\right]=-
\partial_k\frac{\partial}{\partial\psi_\epsilon(x)}
(T_k)_{\epsilon\gamma}
\ee
and
\ba\label{H12}
\big[\cK,\psi_\gamma(x)\big]=-\partial_k\psi_\epsilon(x)(T_k)_{\epsilon\gamma},
\ea
one obtains Dirac equations for real wave functions (with $\partial_0=\partial_t)$
\be\label{H13}
\gamma^\mu \partial_\mu q=0~,~\gamma^\mu \partial_\mu \hat q=0~,~
(\gamma^\mu)^T\partial_\mu\bar q=0.
\ee
We emphasize that these equations follow for arbitrary static states $g_0$ which obey ${\cal K}g_0=0$. 

If needed, we may multiply ${\cal G}_1$ with an appropriate normalization factor such that the wave function obeys
\be\label{57A}
\int_x\sum_\gamma q^2_\gamma(x)=1,
\ee
and $g_1$ has a standard normalization. (No such factor is needed for $g_0=|0\kr$.) We observe that the existence of particles and / or holes depends on the vacuum state $g_0$. For the example $g_0=|0\kr$ no hole states are present, while for $g_0=1$ no particle states are allowed. 

\bigskip\noindent
{\bf  3. Weyl and Majorana spinors}

\medskip
Let us consider $g_0=|0\kr$ where $a_\gamma(x)|0\kr=0$ implies that no hole states exist. There are then only particle states and the propagating degrees of freedom correspond to Majorana fermions. In four dimensions Majorana spinors are equivalent to Weyl spinors \cite{CWMS}. Indeed, we may introduce a complex structure by defining a two-component complex spinor 
\be\label{H14}
\varphi(x)=\left(\begin{array}{c}
\varphi_1(x)\\\varphi_2(x)
\end{array}\right)~,~
\varphi_1=q_1+iq_2~,~\varphi_2=q_3+iq_4.
\ee
The matrices $T_k=\gamma^0\gamma^k$ are compatible with this complex structure. They are translated to the complex Pauli matrices
\be\label{H15}
T_k=\gamma^0\gamma^k\to \tau_k,
\ee
while the operation $\varphi\to i\varphi$ corresponds to the matrix multiplication $q\to \tilde I q$. The operator ${\cal P}_k$ in eq. \eqref{H4} acts on the one-particle states as the usual momentum operator
\be\label{59A}
{\cal P}_k\varphi=-i\partial_k\varphi.
\ee
The Dirac equation reads in the complex basis
\be\label{H16}
\partial_t\varphi=\tau_k\partial_k\varphi~,~i\partial_t\varphi=-\tau_k{\cal P}_k\varphi,
\ee
and the Lorentz generators are given by
\be\label{H17}
\Sigma^{kl}\to-\frac i2\epsilon^{klm}\tau_m~,~
\Sigma^{0k}\to-\frac12\tau_k.
\ee

In contrast, the multiplications with $\gamma^0$ cannot be represented by a multiplication of $\varphi$ with a complex $2\times 2$ matrix, since 
\be\label{18}
q\to \gamma^0 q~~\widehat{=}~~\varphi\to -\tau_2\varphi^*.
\ee
If we express $\varphi^*$ in terms of the two-component complex vector
\be\label{H19}
\chi=E\varphi^*=-i\tau_2\varphi^*=\left(\begin{array}{r}
-q_3+iq_4\\q_1-iq_2\end{array}\right),
\ee
the transformation $q\to\gamma^0 q$ corresponds to 
\be\label{H20}
\varphi\to-i\chi~,~\chi\to -i\varphi.
\ee
We may now introduce the complex four component vector
\be\label{H21}
\Psi_M=\left(\begin{array}{c}
\varphi\\\chi\end{array}\right),
\ee
for which all matrix multiplications $q\to \gamma^\mu q$ can be represented by multiplication with complex matrices of the Clifford algebra,
\ba\label{H22}
&&\gamma^0=\left(\begin{array}{cc}0,&-i\\-i,&0\end{array}\right)~,~
\gamma^k=
\left(\begin{array}{cc}0,&-i\tau_k\\i\tau_k,&0\end{array}\right),\\
&&\Sigma^{0k}=-\frac12\left(\begin{array}{cc}
\tau_k,&0\\0,&-\tau_k
\end{array}\right)~,~
\Sigma^{kl}=-\frac i2\epsilon^{klm}
\left(\begin{array}{cc}
\tau_m,&0\\0,&\tau_m
\end{array}\right).\nn
\ea
We can also define the matrix
\be\label{H23}
\bar\gamma=-i\gamma^0\gamma^1\gamma^2\gamma^3=
\left(\begin{array}{cc}1,&0\\0,&-1\end{array}\right),
\ee
for which $\varphi$ and $\chi$ are eigenvectors with eigenvalues $\pm 1$. (Often $\bar\gamma$ is denoted as $\gamma^5$.) The transformation $q\to\tilde I q$ acts on $\psi_M$ as $\psi_M\to i\bar\gamma \psi_M$. Thus the representation \eqref{H23}, $\bar\gamma=-i\tilde I$, is consistent with the complex structure. 

Since $\chi$ is not independent of $\varphi$ the spinor $\Psi_M$ obeys the Majorana constraint \cite{CWMS}.
\be\label{H24}
B^{-1}\Psi_M^*=\Psi_M~,~B=B^{-1}=-\gamma^2=
\left(\begin{array}{cccc}
0&0&0&1\\0&0&-1&0\\0&-1&0&0\\1&0&0&0
\end{array}\right),
\ee
where $B\gamma^\mu B^{-1}=(\gamma^\mu)^*$. The real Dirac matrices $\gamma^\mu_{(M)}$ in the ``Majorana representation'' \eqref{F7} and the complex Dirac matrices $\gamma^\mu_{(W)}$ in the ``Weyl representation'' \eqref{H22} are related by a similarity transformation,
\be\label{67A}
q=\frac{1}{\sqrt{2}}A\Psi_M~,~\gamma^\mu_{(W)}=A^{-1}\gamma^\mu_{(M)}A,
\ee
with
\be\label{67B}
A=\frac{1}{\sqrt{2}}
\left(\begin{array}{cccc}
1,&0,&,0,&1\\-i,&0,&0,&i\\0,&1,&-1,&0\\0,&-i,&-i,&0
\end{array}\right),
\ee
and
\ba\label{H25}
A^{-1}=\frac{1}{\sqrt{2}}\left(\begin{array}{cccc}
1,&i,&0,&0\\0,&0,&1,&i\\0,&0,&-1,&i\\1&-i,&0,&0
\end{array}\right)=A^\dagger.
\ea

\bigskip\noindent
{\bf   4. Weyl particles}

\medskip
The propagating degrees of freedom correspond to the solutions of the evolution equation. They are most simply discussed in terms of the complex equation \eqref{H16}. We can perform a Fourier transform
\be\label{H26}
\varphi(t,x)=\int_p\tilde \varphi(t,p)e^{ipx}=\int\frac{d^3p}{(2\pi)^3}\tilde\varphi
(t,p)e^{ipx},
\ee
such that the evolution equation becomes diagonal in momentum space,
\be\label{H27}
i\partial_t\tilde\varphi(t,p)=-p_k\tau_k\tilde\varphi(t,p).
\ee
The general solution of eq. \eqref{H16} obeys
\be\label{H28}
\tilde\varphi(t,p)=\exp(ip_k\tau_kt)\tilde \varphi(p),
\ee
with arbitrary complex two-component vectors $\tilde\varphi(p)$. 

We may consider modes with a fixed momentum $p$. (As familiar in quantum mechanics, these modes are not normalizable for infinite volume and may be considered as limiting cases of normalizable wave packets.) For every given $p\neq 0$ we can define 
\be\label{H29}
H(p)=\frac{p_k\tau_k}{|p|}~,~H^2(p)=1~,~p_kp_k=|p|^2,
\ee
such that $H(p)$ has eigenvalues $\pm 1$. Decomposing $\tilde\varphi(p)$ according to the eigenvalues of $H(p)$,
\be\label{30}
H(p)\tilde\varphi_\pm(p)=\mp\tilde\varphi_\pm(p)
\ee
we obtain
\be\label{H31}
\tilde\varphi(t,p)=\exp
\big(-i\omega_+(p)t\big)\tilde\varphi_+(p)+\exp\big(-i\omega_-(p)t\big)
\tilde\varphi_-(p),
\ee
with dispersion relation
\be\label{H32}
\omega_\pm(p)=\pm|p|.
\ee

We observe positive and negative energies $\omega_\pm(p)$. For the complex conjugate of $\varphi(x)$ or $\tilde\varphi(p)$ the sign of $\omega$ is reversed. In a four-component representation \eqref{H21} we can keep modes with positive energy $\omega>0$ as the independent modes, and relate the other components to them by eq. \eqref{H24}. Thus $\varphi$ accounts for one mode $\tilde\varphi_+$ with negative helicity, $H\tilde\varphi_+=-\tilde\varphi_+$, and $\chi$ has a propagating mode $\tilde \chi_-$ with positive helicity, $H\tilde\chi_-=\tilde\chi_-$. We may associate $\varphi$ with a left-handed neutrino or electron, and $\chi$ with a right-handed anti-neutrino or positron. The matrix $\bar\gamma$ \eqref{H23} projects on the left- and right-handed components,
\be\label{78A}
\varphi=\frac12(1+\bar\gamma)\Psi_M~,~\chi=\frac12(1-\bar\gamma)\Psi_M.
\ee
We observe that $\tilde \varphi(p)$ involves for every $t$ two complex or four real functions of momentum. This accounts for two independent propagating charged particle states. The specific state of the propagating modes is determined by the initial data $\tilde \varphi_+(t_0)$ and $\tilde\varphi_-(t_0)$ (or equivalently $\tilde\chi_-(t_0)$). 

We conclude that a particular propagating one-particle state can be characterized by the two complex functions $\tilde\varphi_+(p)$ and $\tilde\varphi_-(p)$, corresponding to positive and negative helicity of the particle. Then $\varphi(t,x)$ and $q_\gamma(t,x)$ can be computed according to eqs. \eqref{H26}, \eqref{H14}. This fixes the Grassmann element $g_1(t)$ by eq. \eqref{H7a} and therefore the coefficients $q_\tau(t)$. For the vacuum $g_0=|0\kr$ the classical wave function $q_\tau(t)$ differs from zero only for those bit chains $\tau=[n_\gamma(x)]$ for which precisely one bit is occupied. The probabilities of the corresponding classical ensemble, $p_\tau(t)=q^2_\tau(t)$, are given for each $\tau$ for which the occupied bit is located at position $x$ and of type $\gamma$ by the simple relation $p_\tau(t)=q^2_\gamma(t,x)$. This gives a direct realization of a one-particle state in a relativistic quantum field theory by a classical statistical ensemble with time evolution \eqref{G14}.

As a concrete example we consider $\tilde\varphi_-(p)$ nonvanishing only for a single momentum $p=(0,0,p_3),p_3>0$. This implies
\be\label{156A}
\tilde\varphi_1(t,p)\sim \exp \{i p_3 t\}\delta\big( p-(0,0,p_3)\big)~,~\tilde\varphi_2(t,p)=0,
\ee
and therefore
\ba\label{158A-1}
\varphi_1(t,x)=L^{-3/2}\exp \{ ip_3(t+x_3)\},
\ea
and
\ba\label{156B}
q_1(t,x)&=&L^{-3/2}\cos \{p_3(t+x_3)\},\nn\\
q_2(t,x)&=&L^{-3/2}\sin\{p_3(t+x_3)\},\nn\\
q_3(t,x)&=&q_4(t,x)=0.
\ea
The probabilities $p(x,\gamma)$ for the Ising model are given by
\ba\label{156C}
p(x,1)&=&L^{-3}\cos^2\{p_3(t+x_3)\},\nn\\
p(x,2)&=&L^{-3}\sin^2\{p_3(t+x_3)\},
\ea
and obey
\be\label{156D}
p(x,1)+p(x,2)=L^{-3}~,~p(x,3)=p(x,4)=0.
\ee
With equal probability a bit is occupied for every point $x$, and only the relative probabilities for the occupation of a bit of species $\gamma=1$ or $\gamma=2$ oscillates in time and in the coordinate $x_3$. A state with momentum $(p_1,0,0)$, $p_1>0$, is realized by 
\be\label{156E}
\varphi_1(t,x)=\varphi_2(t,x)=\frac{1}{\sqrt{2}}L^{-3/2}\exp \{i p_1(t+x_1)\}
\ee
or
\ba\label{156F}
p(x,1)&=&p(x,3)=\frac{1}{2}L^{-3}\cos^2\{p_1(t+x_1)\},\nn\\
p(x,2)&=&p(x,4)=\frac{1}{2}L^{-3}\sin^2\{p_1(t+x_1)\}.
\ea

\bigskip\noindent
{\bf   5. Multi-fermion states}

\medskip
States with arbitrary $n$ describe systems of $n$ fermions. This is not surprising in view of our translation of the classical statistical ensemble to a Grassmann functional integral. For $g_0=|0\kr$ there are only $n$-particle states, and no hole states. They can be constructed by applying $n$ creation operators $a^\dagger_\gamma(x)$ on the vacuum. For example, the two-fermion state obeys
\be\label{H33}
g_2(t)=\frac{1}{\sqrt{2}}\int_{x,y}q_{\gamma\epsilon}(t,x,y)a^\dagger_\gamma(x)
a^\dagger_\epsilon(y)g_0.
\ee
Due to the anticommutation relation
\be\label{H34}
\big\{a^\dagger_\gamma(x)~,~a^\dagger_\epsilon(y)\big\}=
\left\{\frac{\partial}{\partial\psi_\gamma(x)}~,~
\frac{\partial}{\partial\psi_\epsilon(y)}\right\}=0
\ee
the two-particle wave function is antisymmetric, as appropriate for fermions
\be\label{H35}
q_{\gamma\epsilon}(t,x,y)=-q_{\epsilon\gamma}(t,y,x).
\ee
The Grassmann elements $g_2$ describe states with two Weyl spinors. A particular class of states can be constructed as products of appropriately normalized one particle states $q^{(1)},q^{(2)}$
\be\label{H36}
q_{\gamma\epsilon}(x,y)=q^{(1)}_\gamma(x)q^{(2)}_\epsilon(y)-q^{(1)}_\epsilon(y)
q^{(2)}_\gamma(x).
\ee
General two particle states are superpositions of such states. 

It is instructive to specify the probabilities for the Ising model that correspond to one particle with momentum $(0,0,p_3)$ and another with momentum $(p_1,0,0)$. The probabilities $p_\tau$ are now nonvanishing precisely when two bits are occupied, and may be denoted by $p(x,\gamma,y,\delta)=p(y,\delta,x,\gamma)=p_{\gamma\delta}(x,y)$ for the state where the occupied bits are $(x,\gamma)$ and $(y,\delta)$. For the single particle wave functions \eqref{156B}, \eqref{156E} eq. \eqref{H36} yields 
\ba\label{160A}
&&p(x,1,y,1)=\frac14 L^{-6}\big[\cos^2\{p_3(t+x_3)\}\cos^2\{p_1(t+y_1)\}\nn\\
&&+\cos^2\{p_1(t+x_1)\}\cos^2\{p_3(t+y_3)\}\nn\\
&&-2\cos\{p_3(t+x_3)\}\cos \{p_1(t+x_1)\}\nn\\
&&\times\cos \{p_3(t+y_3)\}\cos\{p_1(t+y_1)\}\big].
\ea
The first two terms are not distinguished and represent the product of the probabilities $p(x,1)$ \eqref{156C} and $p(y,1)$ \eqref{156F} written in a symmetrized form. The last term in eq. \eqref{160A} reflects the characteristic interference for quantum particles, with a negative sign as appropriate for fermions. Similarly, one has
\ba\label{160B}
&&p_{12}(x,y)=\frac14 L^{-6}\big[\cos^2\{p_3(t+x_3)\}\sin^2\{p_1(t+y_1)\}\nn\\
&&+\sin^2\{p_3(t+y_3)\}\cos^2\{p_1(t+x_1\}\nn\\
&&-2\cos\{p_3(t+x_3)\}\cos\{p_1(t+x_1)\}\nn\\
&&\times\sin \{p_3(t+y_3)\}\sin \{p_1(t+y_1)\}\big].
\ea
while $p_{22}(x,y)$ and $p_{21}(x,y)$ obtain from $p_{11}(x,y)$ and $p_{12}(x,y)$ by exchanging $\cos\leftrightarrow\sin$, and we observe
\be\label{160C}
\sum_{\gamma,\delta=1,2}p_{\gamma\delta}(x,y)=
\frac{1}{2L^6}\big[1-\cos\{p_3(x_3-y_3)-p_1(x_1-y_1)\}\big].
\ee
One also finds
\be\label{160D}
p_{13}(x,y)+p_{23}(x,y)+p_{14}(x,y)+p_{24}(x,y)=\frac{1}{2L^6},
\ee
and $p_{33}(x,y)=p_{44}(x,y)=p_{34}(x,y)=0$. Due to the interference term $\sim\cos$ in eq. \eqref{160C} the sum of the probabilities over all species for two given positions $(x,y),\bar p(x,y)=\sum_{\gamma\delta}p_{\gamma\delta}(x,y)$ now oscillates in dependence on $x-y$ around a mean value $L^{-6}$. Summing over positions $x$ or $y$ the interference term vanishes, $\bar p(x)=\sum_y\bar p(x,y)=L^{-3}$, such that the probability $\bar p(x)$ to find a particle of an arbitrary species at $x$ is uniform. 

One may wonder why the interference term is necessary. If we drop the interference term in eqs. \eqref{160A}, \eqref{160B}, \eqref{160C} we obtain a perfectly valid probability distribution. However, it does not obey the simple time evolution law \eqref{P35} or \eqref{89A}. Interference effects are a necessary consequence of this law. It is not obvious at all if interesting and nevertheless reasonably simple evolution laws for Ising type systems exist that avoid interference effects. In any case, interference is a very natural consequence of the simple evolution law \eqref{89A}. We note that due to interference the probabilities of finding a particle of any species at $x$ and another at $y$ are correlated. For our example, these correlations persist for arbitrary large separations $(x-y)$. If we replace the momentum eigenstates by wave packets with finite local extension one will find interference and correlations only for regions where the two one-particle wave packets overlap.

\section{Symmetries}
\label{Symmetries}

{\bf 1. Continuous symmetries}

\medskip\noindent
Besides Lorentz symmetry the action \eqref{F1} is also invariant under global $SO(2)$ rotations
\ba\label{71A}
\psi'_1&=&\cos\alpha~\psi_1-\sin\alpha~\psi_2~,~\psi'_2
=\sin\alpha~\psi_1+\cos\alpha~\psi_2,\nn\\
\psi'_3&=&\cos\alpha~\psi_3-\sin\alpha~\psi_4~,~\psi'_4
=\sin\alpha~\psi_3+\cos\alpha~\psi_4.\nn\\
\ea
This is easily seen from the infinitesimal transformation
\be\label{71B}
\delta\psi_\gamma=\alpha\tilde I_{\gamma\delta}\psi_\delta
\ee
and the relations $\tilde I^2=-1,~\tilde I^T=-\tilde I,~[\tilde I,T_k]=0$. These rotations carry over, to the one-particle and one-hole wave functions $q_\gamma$ and $\hat q_\gamma$. 

Correspondingly, the complex two component wave functions $\varphi$ and $\chi$ transform as 
\be\label{71C}
\varphi'=e^{i\alpha}\varphi~,~\chi'=e^{-i\alpha}\chi.
\ee
The $SO(2)$ rotations are now realized as $U(1)$ phase rotations. If we define for a general complex field $\eta$ the charge $\bar Q$ by the transformation 
\be\label{71D}
\eta'=e^{i\alpha \bar Q}\eta
\ee
we infer that $\varphi$ carries charge $\bar Q=1$, while $\chi$ has opposite charge $\bar Q=-1$. If $\varphi$ describes degrees of freedom of an electron, $\chi$ describes the corresponding ones for a positron. We note that charge eigenstates exist only in connection with a complex structure. The real wave function $q_\gamma$ can be encoded both in $\varphi$ and $\chi$ and may therefore describe degrees of freedom with opposite charge. 

\bigskip\noindent
{\bf   2. Discrete symmetries}

\medskip
The action \eqref{F1} or \eqref{N1}, \eqref{C.1}, \eqref{C.4} is further invariant with respect to discrete symmetries. Among them a parity type reflection maps
\be\label{85Aa}
\psi(x)\to\bar{\cal P}\psi(x),
\ee
with
\ba\label{85Ba}
\big (\bar{\cal P}\psi(x)\big)_\gamma&=&(\gamma^0)_{\gamma\delta}\psi_\delta(-x),
\ea
and we note $\bar{\cal P}^2=-1$. We will associate $\bar{\cal P}$ with a CP-transformation since it is realized for Majorana spinors which we may take as eigenstates of charge conjugation $C,C\psi_\gamma=\psi_\gamma$. The action \eqref{F1} is further invariant under the discrete transformation $\psi\to\tilde I\psi$. The continuum limit is therefore also invariant under a parity type transformation where $\bar{\cal P}$ is replaced by $\tilde I\bar {\cal P}$. 

The implementation of time reflection is more subtle. It is realized by a map that involves Grassmann variables at different times. In a discrete setting for time it acts as
\ba\label{166A}
\psi(t,x)\to&\bar T&\psi(t,x),\nn\\
\bar T\psi(t,x)=&\hat T&\psi(-t,x)\text{ for }t \text{ even},\nn\\
\bar T\psi(t,x)=-&\hat T&\psi(-t,x)\text{ for } t\text{ odd},
\ea
where the matrix $\hat T$ obeys
\ba\label{17xx}
\hat T=\tilde I\gamma^0~,~\hat T^T\hat T=1~,~ \hat T^TT_k\hat T=-T_k.
\ea
The different transformation of spinors at even or odd time points for the discrete formulation is crucial, since otherwise the invariance of $\sum_{t,x}\psi_\gamma (t)\psi_\gamma(t+\epsilon)$ could not be realized by an orthogonal matrix $\hat T$. The action \eqref{N1}, \eqref{C.1}, \eqref{C.4} is invariant under the time reflection \eqref{166A}. The relative minus sign in the transformation of the two spinors in the action \eqref{F1} - one at even, the other at odd $t$ - has to be remembered for the invariance of the continuum action. We can combine the transformations \eqref{85Ba}, \eqref{166A} to a PT transformation
\ba\label{166B}
\psi(t,x)\to&\tilde I&\psi(-t,-x)\text{ for } t\text{ even}\nn\\
\psi(t,x)\to-&\tilde I&\psi (-t,-x)\text{ for }t\text{ odd}.
\ea
For Majorana spinors this plays the role of CPT symmetry.

\section{Complex structure}
\label{Complex structure}

We have formulated the classical statistical description of a quantum field theory for Majorana spinors in terms of a real Grassmann algebra. All quantities in the functional integral \eqref{N9} and the action \eqref{F1} or \eqref{N1}, \eqref{C.1}, \eqref{C.4} are real. The introduction of complex Weyl spinors in eq. \eqref{H14} or \eqref{H19} reveals the presence of a complex structure in this real formulation. This will be discussed in more detail in the present section. 

\bigskip\noindent
{\bf   1. Complex Grassmann variables}

\medskip
Complex Grassmann variables may be introduced in analogy to eq. \eqref{H14} 
\be\label{71LA}
\zeta_1=\frac{1}{\sqrt{2}}(\psi_1+i\psi_2)~,~
\zeta_2=\frac{1}{\sqrt{2}}(\psi_3+i\psi_4).
\ee
Together with the complex conjugate Grassmann variables
\be\label{158A}
\zeta^*_1=\frac{1}{\sqrt{2}}(\psi_1-i\psi_2)~,~\zeta^*_2=\frac{1}{\sqrt{2}}
(\psi_3-i\psi_4)
\ee
we have for every $x$ and $t$ four independent Grassmann variables $\zeta_1,\zeta^*_1,\zeta_2,\zeta^*_2$, which replace $\psi_1,\psi_2,\psi_3,\psi_4$. 

A general element of a complex Grassmann algebra can be expanded as
\ba\label{71M}
g_c&=&\sum_{k,l}c_{\alpha_1\dots \alpha_k,\bar\alpha_1\dots\bar\alpha_l}
(x_1\dots x_k,\bar x_1\dots \bar x_l)\nn\\
&&\zeta_{\alpha_1}(x_1)\dots\zeta_{\alpha_k}(x_k)\zeta^*_{\bar\alpha_1}(\bar x_1)\dots
\zeta^*_{\bar\alpha_l}(\bar x_l),
\ea
with complex coefficients $c$. (For a real Grassmann algebra the coefficients $c$ are restricted by $g^*=g$.) From a general complex $g_c$ we can obtain elements of a real Grassmann algebra as
\be\label{87A}
g=\frac12(g_c+g^*_c)~,~g'=-\frac i2(g_c-g^*_c).
\ee

On every factor $\zeta$ the action of $\tilde I$ amounts to a multiplication with $i$. More precisely, we can interpret eq. \eqref{71LA} as a map $\psi\to\zeta[\psi]$ with the property $\zeta[\tilde I\psi]=i\zeta[\psi]$. For the complex conjugate one has $\zeta^*[\tilde I\psi]=-i\zeta^*[\psi]$. For the infinitesimal transformation \eqref{71B} one concludes 
\be\label{71N}
\delta\big [\zeta_{\alpha_1}(x_1)\dots \zeta^*_{\bar\alpha_l}(\bar x_l)\big ]=
i\alpha\bar Q\zeta_{\alpha_1}(x_1)\dots\zeta^*_{\bar\alpha_l}(\bar x_l),
\ee
where the charge $\bar Q$ counts the number of factors $\zeta$ minus the number of factors $\zeta^*$ for a given term in the expansion \eqref{71M}. In other words, products with $\bar Q_+$ factors $\zeta$ and $\bar Q_-$ factors $\zeta^*$ are charge eigenstates with $\bar Q=\bar Q_+-\bar Q_-$. States with a given $\bar Q$ are degenerate since many different choices of $\bar Q_+,\bar Q_-$ lead to the same $\bar Q$. For a real Grassmann algebra we use eq. \eqref{87A}. Every term in $g_c$ with $\bar Q\neq 0$ is accompanied by a term with opposite charge $-\bar{Q}$ in $g^*_c$. Thus the expansion \eqref{71M} of eq.  \eqref{87A} involves ``charge eigenspaces'' with pairs of opposite axial charge, similar to $\psi_\gamma$ in eq. \eqref{71A}. 

\bigskip\noindent
{\bf  2. Complex Grassmann algebra}

\medskip
It will often be convenient to use a complex Grassmann algebra, where the coefficients $c$ in eq. \eqref{71M} are arbitrary and charge eigenstates belong to the Grassmann algebra also for $\bar Q\neq 0$. This can be mapped at the end to a real Grassmann algebra by eq. \eqref{87A}. The action of $\bar Q_\pm$ on $g_c$ is represented as 
\be\label{71O}
\bar Q_+=\int_y\zeta_\alpha(y)\frac{\partial}{\partial\zeta_\alpha(y)}~,~
\bar Q_-=\int_y\zeta^*_\alpha(y)\frac{\partial}{\partial\zeta^*_\alpha(y)},
\ee
with
\ba\label{168Aa}
\frac{\partial}{\partial\zeta_1(y)}&=&\frac{1}{\sqrt{2}}
\left(\frac{\partial}{\partial\psi_1(y)}-i
\frac{\partial}{\partial\psi_2(y)}\right),
\ea
and
\ba\label{71OA}
\frac{\partial}{\partial\zeta^*_1(y)}&=&
\left(\frac{\partial}{\partial\zeta_1(y)}\right)^*\nn\\
&=&\frac{1}{\sqrt{2}}\left(\frac{\partial}{\partial\psi_1(y)}+i
\frac{\partial}{\psi_2(y)}\right),
\ea
and similar for $\partial/\partial\zeta_2(y)$ and $\partial/\partial\zeta^*_2(y)$, with $\psi_{1,2}$ replaced by $\psi_{3,4}$. 

The Grassmann derivatives $\partial/\partial \zeta_\alpha$ obey the standard anticommutation relations
\ba\label{104A}
\left\{\frac{\partial}{\partial\zeta_\alpha(x)},\zeta_\beta(y)\right\}&=&
\delta_{\alpha\beta}\delta(x-y)~,~
\left\{\frac{\partial}{\partial\zeta_\alpha(x)},\zeta^*_\beta(y)\right\}=0.\nn\\
\ea
For $\bar Q=\bar Q_+-\bar Q_-$ eq. \eqref{71O} yields
\be\label{71P}
-i\bar Q=\int_y\frac{\partial}{\partial\psi_\gamma(y)}\tilde I_{\gamma\delta}\psi_\delta(y)
\ee
and we recognize the conserved quantity $B_2$ in eq. \eqref{H3a} with $b_2=1$. On the other hand, the sum yields the conserved quantity ${\cal N}$ via $\bar Q_++\bar Q_-=B-{\cal N}$. 

In the complex basis we can write the time evolution equation in the form
\be\label{71Q}
i\partial_t g_c={\cal H}g_c
\ee
with 
\be\label{71R}
{\cal H}=i\int_x\left[\frac{\partial}{\partial\zeta_\alpha}
(\tau_k\partial_k\zeta)_\alpha+\frac{\partial}{\partial\zeta^*_\alpha}
(\tau^*_k\partial_k\zeta^*)_\alpha\right].
\ee
Both $\bar Q_+$ and $\bar Q_-$ commute with ${\cal H}$. The operators creating one hole or one particle states read
\ba\label{93A}
\hat q_\gamma\psi_\gamma&=&\frac{1}{\sqrt{2}}
(\hat\varphi^*_\alpha\zeta_\alpha+\hat\varphi_\alpha\zeta^*_\alpha),\nn\\
q_\gamma\frac{\partial}{\partial\psi_\gamma}&=&\frac{1}{\sqrt{2}}
\left(\varphi_\alpha\frac{\partial}{\partial\zeta_\alpha}+\varphi^*_\alpha
\frac{\partial}{\partial\zeta^*_\alpha}\right).
\ea

\bigskip\noindent
{\bf   3. Similarity transformation}

\medskip
Defining four component vectors $\psi=(\psi_1,\psi_2,\psi_3,\psi_4)$ and $\zeta_d=(\zeta_1,\zeta_2,\zeta^*_1,\zeta^*_2)=(\zeta,\zeta^*)$ the relations \eqref{71LA}, \eqref{158A} can be written as a similarity transformation
\ba\label{174A}
\psi&=&G\zeta_d~,~\zeta_d=G^{-1}\psi=
G^\dagger\psi,\nn\\
G^\dagger G&=&1~,~\det G=1,
\ea
with
\ba\label{168A}
G&=&\frac{1}{\sqrt{2}}\left(\begin{array}{rrrr}
1,&0,&1,&0\\
-i,&0,&i,&0\\
0,&1,&0,&1\\
0,&-i,&0,&i
\end{array}\right),\nn\\
G^{-1}&=&\frac{1}{\sqrt{2}}
\left(\begin{array}{rrrr}
1,&i,&0,&0\\
0,&0,&1,&i\\
1,&-i,&0,&0\\
0,&0,&1,&-i
\end{array}\right).
\ea
The Grassmann derivatives ${\partial}/{\partial\zeta_d}$ obey the standard anticommutation relations (cf. eq. \eqref{104A}) 
\be\label{168B}
\frac{\partial}{\partial\zeta_d}=G^T\frac{\partial}{\partial\psi}~,~
\left\{\frac{\partial}{\partial\zeta_{d,u}}~,~\zeta_{d,v}\right\}=\delta_{uv}.
\ee
Grassmann bilinears transforms as $\psi^TM\psi=\zeta^T_d M_\zeta\zeta_d$ with $M_\zeta=G^TMG$. In particular, one finds
\be\label{168C}
G^TG=\left(\begin{array}{c}
0,1\\1,0
\end{array}\right)~,~
(T_k)_\zeta=G^T T_kG=
\left(\begin{array}{rl}
0,&\tau^*_k\\\tau_k,&0
\end{array}\right),
\ee
such that the action \eqref{F1} reads
\ba\label{168D}
S&=&\int_{t,x}\big\{\zeta^\dagger(\partial_t-\tau_k\partial_k)\zeta+\zeta^T
(\partial_t-\tau^*_k\partial_k)\zeta^*\big\}\nn\\
&=&2\int_{t,x}\zeta^\dagger(\partial_t-\tau_k\partial_k)\zeta.
\ea
Up to the factor $2$, which may be removed by a rescaling of $\zeta$, this is the action for a free Weyl spinor.

Transferring a linear transformation $\psi'=L\psi$ to $\zeta_d,\zeta'_d=L'\zeta_d$, requires the transformation property $L'=G^{-1}LG=G^\dagger LG$. We note the difference between the transformation property of the kernel of a bilinear $M$ which involves $G^T$, and the transformation of $L$ which involves $G^{-1}=G^\dagger$. In particular, one finds
\be\label{168E}
G^\dagger T_kG=\left(\begin{array}{rl}
\tau_k,&0\\0,&\tau^*_k
\end{array}\right).
\ee
This difference is connected to the additional factor $G^TG$ appearing in the transformation of the bilinear
\be\label{168F}
(G^TG)(G^\dagger T_kG)=(T_k)_\zeta.
\ee
If we define bilinears in terms of
\be\label{168G}
\zeta^\dagger_d=\zeta^T_d(G^TG)=\zeta^T_d{0,1 \choose 1,0}=
\left(\zeta^T_d\right)^*=(\zeta^*,\zeta),
\ee
i.e. by $\zeta^\dagger_d\bar M_\zeta\zeta_d$, then $\bar M_\zeta$ has the same transformation property as $L$. One finds
\be\label{168H}
G^\dagger\tilde IG=
\left(\begin{array}{rl}
i,&0\\0,&-i
\end{array}\right)
\ee
and the generators $\Sigma$ of the Lorentz group \eqref{F13} are transformed to 
\be\label{168I}
G^\dagger\Sigma G=
\left(\begin{array}{rl}
\Sigma',&0\\0,&\Sigma'^*
\end{array}\right)
\ee
with $2\times 2$-matrices $\Sigma'$ given by eq. \eqref{H17}.

More explicitly, the infinitesimal Lorentz transformations of the complex two-component spinors $\zeta,\delta\zeta=-\frac12\epsilon_{mn}\Sigma^{mn}\zeta$, are represented by the complex $2\times 2$  matrices $\Sigma^{mn}$ given by eq. \eqref{H17}. We also observe that  $\tilde\zeta=E\zeta=-i\tau_2\zeta$ transforms as $\delta\tilde\zeta=\frac12\epsilon_{mn}(\Sigma^{mn})^T\tilde\zeta$ such that
\be\label{93B}
\frac12\tilde\zeta^T\zeta=\zeta_1\zeta_2=\frac12\big[\psi_1\psi_3-\psi_2\psi_4+i
(\psi_1\psi_4+\psi_2\psi_3)\big]
\ee
is a Lorentz scalar. The same holds for $\zeta^*_1\zeta^*_2$ such that the bilinears $\psi_1\psi_3-\psi_2\psi_4$ and $\psi_1\psi_4+\psi_2\psi_3$ are separately Lorentz scalars. Also the product
\be\label{93C}
\zeta_1\zeta_2\zeta^*_1\zeta^*_2=\psi_1\psi_2\psi_3\psi_4=
\frac{1}{24}\epsilon^{\alpha\beta\gamma\delta}
\psi_\alpha\psi_\beta\psi_\gamma\psi_\delta
\ee
is a Lorentz scalar. The functional measure obeys
\ba\label{93D}
&&\int d \psi_4 d\psi_3 d\psi_2 d\psi_1=\prod_\gamma d\psi_\gamma\nn\\
&&=\int d\zeta_2 d\zeta_1\int d\zeta^*_2 d\zeta^*_1=
\int d \zeta d\zeta^*.
\ea

\bigskip\noindent
{\bf   4.  Complex structure for real Grassmann algebra}

\medskip
In a real even-dimensional vector space a complex structure is given by the existence of an involution $K$, together with a map $I$, obeying
\be\label{186A}
K^2=1~,~I^2=-1~,~\{K,I\}=0.
\ee
The matrix $K$ has eigenvalues $\pm 1$ and we may denote eigenstates with positive eigenvalues by $v_R$ and those with negative ones by $v_I,Kv_R=v_R, Kv_I=-v_I$. The matrix $I$ is a map  between $v_R$ and $v_I$, implying that the number of independent $v_R$ and $v_I$ are equal. We can choose the $v_I$ such that $Iv_R=v_I,Iv_I=-v_R$ and use these properties for defining a map from the real vectors $v$ to complex vectors $c=v_R+iv_I=c(v)$, with the properties $c(Kv)=\big[c(v)\big]^*$, $c(Iv)=ic(v)$. An example is the map \eqref{H14} acting on $v=\{q_\gamma\}$ with 
\be\label{186B}
K=\tilde K=\left(\begin{array}{rr}
\tau_3,&0\\0,&\tau_3
\end{array}\right)~,~I=\tilde I.
\ee
Complex structures can also be defined for $K$ and $I$ acting only on subspaces.

We may define a complex structure on the space of the Grassmann variables, $v\widehat{=}\psi_\gamma$, using eq. \eqref{186B}. This results in the map $\zeta[\psi]$ given by eq. \eqref{71LA}, with 
\be\label{186C}
\zeta[\tilde K\psi]=\zeta^*[\psi]~,~\zeta[\tilde I\psi]=i\zeta[\psi],
\ee
where $\zeta$ plays the role of $c$. A transformation $\psi\to A\psi$ is compatible with this complex structure if $A$ obeys
\be\label{186D}
[A,\tilde I]=0.
\ee
In this case $A$ can be represented by complex matrix multiplication acting on $\zeta$,
\be\label{186E}
\zeta[A\psi]=\tilde A\zeta[\psi].
\ee
The matrices $T_k$ commute with $\tilde I$ and are therefore compatible with the complex structure
\be\label{186F}
\zeta[T_k\psi]=\tau_k\zeta[\psi].
\ee
Also $\tilde I$ and $\Sigma^{\mn}$ (cf. eq. \eqref{F5}) are compatible with the complex structure, where the action of $\Sigma^{\mn}$ on $\zeta$ is given by eq. \eqref{H17}. The matrices $A$ obeying eq. \eqref{186D} form a group, since the product of two matrices $A_1A_2$ again commutes with $\tilde I$. This product is represented by complex matrix multiplication of $\tilde A_1$ and $\tilde A_2$, $\zeta[A_1A_2\psi]=\tilde A_1\tilde A_2\zeta[\psi]$. In contrast, the matrices $\gamma^\mu$ anticommute with $\tilde I$, cf. eq. \eqref{F12}. They are therefore not compatible with the complex structure \eqref{186B}. (No set of four mutually anticommuting complex $2\times 2$ matrices exists.) One finds
\be\label{186G}
\zeta[\gamma^0\psi]=-\tau_2\zeta^*.
\ee

Eq. \eqref{186B} defines a complex structure for the Grassmann elements involving only one factor of $\psi$. One can extend this complex structure to Grassmann elements with an arbitrary number of factors $\psi$, except for the elements $|0\kr$ and $|1\kr$. The complex conjugation corresponding to eqs. \eqref{71LA}-\eqref{71M} reverses the sign of all basis elements $g_\tau$ which have an odd number of factors $\psi_2$ or $\psi_4$. We will not discuss this issue further. It should be mentioned, however, that many different complex structures can be defined for the Grassmann algebra. For example, one could use for the involution $K$ the conjugation ${\cal C}$.

\section{Particle-hole conjugation}
\label{Particle-holeconjugation}
The time evolution of the probability distribution $\big\{p_\tau(t)\big\}$ for the classical statistical ensemble is invariant under the exchange of occupied and empty bits. For every classical state $\tau=\big[n_\gamma(x)\big]$ we can define the anti-state $\bar\tau=\big[\bar n_\gamma(x)\big]$ with $\bar n_\gamma(x)=1-n_\gamma(x)$. The particle-hole conjugation map $\tau\to\bar\tau$ transforms occupied bits to empty bits and vice versa. The particle number of a state $\tau$, given by $N_\tau=\sum_\gamma\sum_xn_\gamma(x)=m$, corresponds for the anti-state to $N_{\bar\tau}=\sum_\gamma\sum_x\big(1-n_\gamma(x)\big)=B-m$. The particle-hole conjugate of a probability distribution $\{p_\tau\}$ exchanges the role of $\tau$ and $\bar\tau$, i.e. $C\big(\{p_\tau\}\big)=\{p^c_\tau\}$, with $p^c_\tau=p_{\bar\tau}$.

\bigskip\noindent
{\bf   1. Particle-hole conjugate wave functions}

\medskip
On the level of the Grassmann algebra the particle-hole conjugation maps Grassmann elements $g_m\leftrightarrow \pm g_{B-m}$. Two probability distributions or associated Grassmann elements which are related by particle-hole conjugation obey the same evolution equation. This is expressed by the invariance of the evolution operator $\cK$ \eqref{P46} under the exchange of $\psi_\gamma(x)$ and $\partial/\partial\psi_\gamma(x)$, and in the identical evolution equations for $q$ and 
$\hat q$ in eq. \eqref{H13}. 

The particle-hole conjugation of Grassmann elements $g(\psi)$ can be realized by the operation ${\cal C}$ which maps $\psi_\gamma(x)$ factors into factors $1$, and inserts $\eta_\gamma\psi_\gamma(x)$ for the places $(\gamma,x)$ where no such factor was present. In sect. \ref{Probabilitydistribution} we have defined the conjugate Grassmann element $\tilde g_\tau$ (no sum over $\tau$)
\be\label{106A}
\tilde g_\tau g_\tau=|0\kr~,~{\cal C} g_\tau=\sigma_\tau\tilde g_\tau~,~
{\cal C}\tilde g_\tau=\sigma_\tau g_\tau.
\ee

We may define the operation $\tilde K,g\to \tilde K g$, which multiplies each factor $\psi_\gamma(x)$ by $\eta_\gamma$, cf. eq. \eqref{186B}. This can be used to write
\be\label{188A}
{\cal C}=\tilde K{\cal C}_{ph}~,~{\cal C}^2_{ph}=1,
\ee
where ${\cal C}_{ph}$ replaces $\psi_\gamma(x)\leftrightarrow 1$ without additional sign factors. (The operation $\tilde K$ could be used for defining a complex conjugation, $\tilde K^2=1$.) The operation ${\cal C}_{ph}$ is another realization of the particle-hole conjugation of the probability distribution $\{p_\tau\}$, since $\cal C$ and ${\cal C}_{ph}$ only differ by signs of the components of the wave function $q_\tau$. 

The particle-hole conjugation ${\cal C}_{ph}$ obeys the relations
\be\label{188B}
{\cal C}_{ph}\psi_\gamma 
(x){\cal C}_{ph}g=\frac{\partial}{\partial\psi_\gamma(x)}g~,~
{\cal C}_{ph}\frac{\partial}{\partial\psi_\gamma(x)}
{\cal C}_{ph}g=\psi_\gamma(x)g.
\ee
For an arbitrary Grassmann operator ${\cal A}$ we define the particle-hole conjugate operator ${\cal A}^c$ as 
\be\label{188C}
{\cal A}^c={\cal C}_{ph}{\cal A}{\cal C}_{ph}.
\ee
Since we can write ${\cal A}$ as a sequence of suitable operators $\psi_\gamma(x)$ and $\partial/\partial\psi_\gamma(x)$ we conclude from eq. \eqref{188B} that the operator ${\cal A}^c$ obtains from ${\cal A}$ by exchanging $\psi_\gamma(x)\leftrightarrow\partial/\partial\psi_\gamma(x)$. We observe 
\ba\label{106C}
{\cal C}_{ph}\left(\frac{\partial}{\partial\psi_\gamma(x)}|0\kr\right)&=&
{\cal C}_{ph}\big(-\tilde\psi_\gamma(x)\big)=\psi_\gamma(x),\nn\\
{\cal C}_{ph}\psi_\gamma(x)&=&-\tilde \psi_\gamma(x).
\ea

The conjugate of the totally empty possible vacuum state $g_0=|0\kr$ is given by the totally occupied sequence $g_0=|1\kr$, represented by the Grassmann element $1$, with $\cN g_0=B g_0=m_0g_0$. Correspondingly, the particle wave functions $q_\gamma(x)$ for an expansion around $|0\kr$ are mapped by $C_{ph}$ into the hole wave functions $\hat q_\gamma(x)$ for an expansion around $|1\kr$, replacing the creation operators $a^\dagger_\gamma$ by annihilation operators $a_\gamma$. The discussion of hole states is then completely parallel to the particle states. We observe that the particle-hole conjugate states can also be described in the expansion around the vacuum $|0\kr$. For example, the one hole state $a_\gamma(x)|1\kr$ obtains by applying $B-1$ creation operators on $|0\kr$. The direct construction by applying the particle-hole conjugation to a one particle state is a simple way of constructing these states. In the following we will use a generalized definition of one-hole states as the particle-hole conjugate of the one-particle states. These states are always contained in the Grassmann elements $g$. For a particle-hole symmetric vacuum $g_0$ this generalized notion coincides with the definition \eqref{H8a}. If ${\cal C}_{ph}g_0=\tilde g_0$ does not coincide with $g_0$ we replace eq. \eqref{H8a} by
\be\label{106X}
g_{-1}(t)=\int_x\hat q_\gamma(x)a_\gamma(x)\tilde g_0.
\ee
For $\cN g_0=m_0g_0,\cN\tilde g_0=(B-m_0)\tilde g_0$ one has now
\be\label{106Y}
\cN g_{-1}=(B-m_0-1) g_{-1}.
\ee

Two states which are related by particle-hole conjugation have the same Lorentz transformation properties. This follows from the observation that for a Lorentz invariant $g_0$ also $\tilde g_0$ is Lorentz invariant,together with the transformation properties of the creation and annihilation operators. In particular, the one-hole wave function $\hat q_\gamma(x)$ transforms in the same way as the one-particle wave function $q_\gamma(x)$. 

We may represent the one-hole wave function by a complex two-component spinor 
\be\label{83A}
\hat\varphi(x)=\left(\begin{array}{c}
\hat q_1(x)+i\hat q_2(x)\\
\hat q_3(x)+i\hat q_4(x)
\end{array}\right)
\ee
and use similar to eq. \eqref{H19}
\be\label{K2}
\hat \chi=E\hat\varphi^*=\left(\begin{array}{c}
-\hat q_3+i\hat q_4\\\hat q_1-i\hat q_2
\end{array}\right).
\ee
In the representation by the complex two-component spinor $\hat\chi$ one finds
\ba\label{K3}
T_k&=&\gamma^0\gamma^k=-2\Sigma^{0k}\to -\tau_k,\nn\\
\Sigma^{kl}&=&-\frac14[T_k,T_l]\to -\frac i2\epsilon^{klm}\tau_m.
\ea
In this representation the particle-hole conjugation ${\cal C}_{ph}$ maps a left handed particle with charge $\bar Q=1$ into a right handed hole with charge $\bar Q=-1$ 

We could combine the one-particle and one-hole states and define a complex four component spinor $\Psi_{ph}$ similar to eq. \eqref{H21}, 
\be\label{86}
\Psi_{ph}={\varphi\choose\hat\chi}.
\ee
The Dirac matrices acting on $\Psi_{ph}$ have the representation given by eqs. \eqref{H22}, \eqref{H23}. The Majorana constraint \eqref{H24} does no longer hold, however, since $\varphi$ and $\hat\chi$ are independent. The spinor $\Psi_{ph}$ describes now a type of Dirac spinor which contains degrees of freedom for both particles and holes, corresponding to the eight real functions $q_\gamma(x)$ and $\hat q_\gamma(x)$. We may denote the particle-hole conjugate Dirac spinor by
\be\label{87}
\Psi_{hp}=B^{-1}\Psi_{ph}^*.
\ee
As it should be, it obtains from $\Psi_{ph}$ by $q_\gamma\to \hat q_\gamma$. 

One may be tempted to identify $\hat\varphi$ with the left-handed positron and $\hat\chi$ with the right-handed electron. However, the spinors $\varphi$ and $\hat\varphi$ have both $\bar Q=1$, while $\chi$ and $\hat\chi$ carry both $\bar Q=-1$. The upper two and lower two components of $\Psi_{ph}$ carry a different charge $\bar Q$, such that $\hat\chi$ cannot be identified with the right handed electron. (This should carry the same charge as the left-handed one.) The Dirac spinor \eqref{86} describes a left handed particle and a right handed hole with opposite charge, including the complex conjugate states. This amounts to two independent Majorana-spinors, or equivalently to two left-handed Weyl spinors $\varphi$ and $\hat\varphi$ with identical  charge. If we want to describe the usual electron and positron we would need $\varphi$ and $\hat\chi$ to carry the same charge.

\bigskip\noindent
{\bf  2. Particle-hole symmetric vacua}

\medskip
An interesting situation arises for a vacuum state $g_0$ which is invariant under the particle-hole conjugation,
\be\label{85A}
{\cal C}_{ph}g_0=g_0.
\ee
An example is the linear superposition
\be\label{85B}
g_0=\frac{1}{\sqrt{2}}(|0\rangle+|1\rangle)
\ee
and we will discuss other examples below. For a particle-hole symmetric vacuum obeying eq. \eqref{85A} both particle and hole states are present in the expansion around $g_0$, applying either creation or annihilation operators. (One has to adapt the normalization of eqs. \eqref{H7a}, \eqref{H8a}.) The particular vacuum state \eqref{85B} has not a fixed particle number, since $\cN|0\kr=0,\cN|1\kr=B|1\kr$. It corresponds to a classical probability distribution with probability $1/2$ if either all $n_\tau=1$ or if all $n_\tau=0$. 

A perhaps more interesting candidate for a particle-hole symmetric vacuum is a half-filled state with $N=B/2$. A possible candidate is given in the complex representation \eqref{71LA}, \eqref{93B} by
\ba\label{111A}
g_{c,0}=\prod_x\big\{\zeta_1(x)\zeta_2(x)\big\}~,~
g_0=\frac{1}{\sqrt{2}}(g_{c,0}+g^*_{c,0}).
\ea
This state is Lorentz invariant and obeys
\ba\label{111B}
{\cal C}_{ph}g_{c,0}&=&g^*_{c,0}~,~\cN g_{c,0}=\frac{B}{2}g_{c,0},\nn\\
\bar Q g_{c,0}&=&\frac B2 g_{c,0}~,~g^*_{c,0}g_{c,0}=|0\kr,
\ea
and correspondingly 
${\cal C}_{ph}g^*_{c,0}=g_{c,0},~ \cN g^*_{c,0}=\frac B2 g^*_{c,0},~\bar Q g^*_{c,0}=-
\frac B2 g^*_{c,0}$, such that $g_0$ is invariant under particle-hole conjugation and normalized, $g^2_0=|0\kr$. 

In order to show that $g_0$ is a static state we use discrete time steps such that $g(t+\epsilon)$ obtains from $g(t)$ by the replacement
\be\label{111C}
\zeta(x)=\left(\begin{array}{c}
\zeta_1(x)\\\zeta_2(x)
\end{array}\right)\to A\zeta(x)-\epsilon\tau_k\partial_k\zeta(x).
\ee
If $g_{c,0}(t)$ has the form \eqref{111A} one infers from eq. \eqref{111C} 
\ba\label{111D}
&&g_{c,0}(t+\epsilon)=\prod_x
\Big\{A^2\zeta_1(x)\zeta_2(x)+\frac{A\epsilon}{2\Delta}\big[\zeta_2(x)
(\tau_k)_{1\alpha}\nn\\
&&-\zeta_1(x)(\tau_k)_{2\alpha}\big]\big[\zeta_\alpha(x+\Delta_k)-\zeta_\alpha
(x-\Delta_k)\big]\nn\\
&&+\frac{\epsilon^2}{4\Delta^2}
(\tau_k)_{1\alpha}(\tau_l)_{2\beta}
\big[\zeta_\alpha(x+\Delta_k)-\zeta_\alpha(x-\Delta_k)\big]\nn\\
&&\times\big[\zeta_\beta(x+\Delta_l)-\zeta_\beta(x-\Delta_l)\big]\Big\}.
\ea
The combinatorics of this expression looks at first sight complicated. However, eq. \eqref{111D} contains precisely $B/2$ factors $\zeta$ which have to be distributed on $B/4$ sites $x$. For every $x$ we can have at most two factors of $\zeta$, since $\zeta^2_1(x)=\zeta^2_2(x)=0$. The only nonvanishing contribution must therefore have precisely one factor of $\zeta_1$ and one factor of $\zeta_2$ for every $x$, and we conclude
\be\label{111E}
g_{c,0}(t+\epsilon)=\kappa g_{c,0}(t).
\ee
We next show that the proportionality constant $\kappa$ is real, $\kappa^*=\kappa$. Indeed, an imaginary part of this coefficient could only arise from terms in eq. \eqref{111D} which involve an odd number of $\tau_2$-factors. On the other hand, all terms involving an odd number of factors $\tau_k$ for any $k$ must vanish. Such terms have an odd number of $\zeta$-factors displaced by one lattice unit $\pm \Delta_k$, and this is not compatible with having two $\zeta$-factors for every $x$. Finally, we have seen that the time evolution preserves the normalization of $g(t)$ and conclude $\kappa=1$. Thus $g_0$ in eq. \eqref{111A} is indeed a static state, ${\cal K} g_0=0$. In consequence of eq. \eqref{93B} the state $g_0$ is Lorentz-invariant and we can define one particle and one hole states by applying annihilation and creation operators. We notice, however, that $g_0$ is not a state with zero axial charge $\bar Q$. 

The classical ensemble which corresponds to the half-filled vacuum \eqref{111A} has equal static probabilities for $4^{L^3}/2$ Ising states, while the other Ising states have zero probability. The states with nonzero probability can be characterized as follows: (i) On every lattice site two bits are occupied and two bits are empty. (ii) The occupied bits on every site are in one of the allowed combinations of species $(1,3),(1,4),(2,3),(2,4)$. This follows from the property
\be\label{XXA}
\zeta_1(x)\zeta^*_1(x)g_0=\zeta_2(x)\zeta^*_2(x)g_0=0,
\ee
which forbids the species combinations $(1,2)$ and $(3,4)$. (iii) Only $K$-even states have novanishing probability. The $K$-even states have an even number of occupied bits of the species $2$ and $4$, while $K$-odd states have an odd number of occupied bits of this type. This follows from the map $K$ which maps $g_{c,0}\to g^*_{c,0}$ and reverses the sign of all terms in $g$ with an odd number of Grassmann variables $\psi_2$ or $\psi_4$. 

\bigskip\noindent
{\bf 3. Particle and hole states for the half-filled vacuum}

\medskip
The normalized one particle state in an expansion around $g_0$ \eqref{111A} is given by the Grassmann element
\ba\label{111F}
g_1(t)&=&\int_x q_\gamma(t,x) a^\dagger_\gamma(x)(g_{c,0}+g^*_{c,0})\\
&=&\frac{1}{\sqrt{2}}\int_x
\left\{\varphi_\alpha(x)\frac{\partial}{\partial\zeta_\alpha(x)}
g_{c,0}+\varphi^*_\alpha(x)
\frac{\partial}{\partial\zeta^*_\alpha(x)}g^*_{c,0}\right\}.\nn
\ea
Similarly, the one hole state reads
\ba\label{111G}
g_{-1}(t)&=&\int_x\hat q_\gamma(t,x)a_\gamma(x)
(g_{c,0}+g^*_{c,0})\\
&=&\frac{1}{\sqrt{2}}\int_x
\big\{\hat\varphi_\alpha(x)\zeta^*_\alpha (x) g_{c,0}+\hat\varphi^*_\alpha (x)
\zeta_\alpha (x) g^*_{c,0}\big\},\nn
\ea
and we infer that $\varphi$ and $\hat\varphi$ both carry axial charge $\bar Q=1-B/2$. For all $t$ one finds the relation
\be\label{111H}
g_{-1}(t)g_1(t)=\frac12
\int_x\big(\varphi^*_\alpha(x)\hat\varphi_\alpha(x)+\hat\varphi^*_\alpha(x)
\varphi_\alpha(x)\big)|0\kr.
\ee
For $\hat\varphi_\alpha(x)=\varphi_\alpha(x)$ the one hole state $g_{-1}$ is the particle-hole conjugate state of $g_1$, such that with the normalization of the wave function
\be\label{111I}
\int_x\varphi^*_\alpha(x)\varphi_\alpha(x)=1
\ee
one has
\be\label{111J}
\int {\cal D}\psi g_{-1}(t) g_1(t)=1.
\ee

The one-particle or one-hole states \eqref{111F} and \eqref{111G} differ from the states $\cG_1|0\kr$ and $\cG_{-1}|1\kr$. By now we have already found four Grassmann elements for which the wave function $q_\gamma(x)$ or $\hat q_\gamma(x)$ obeys the Dirac equation \eqref{57A}. They are all eigenstates of $\cN$ with eigenvalues $1,B-1,~B/2+1$ and $B/2-1$. All these states are described by the functional integral \eqref{N9} for a suitable choice of the boundary term $g_{in}$. They correspond to different time sequences of classical probability distributions $p_\tau(t)$. The question to what extent these different states have to be considered as different particles or should be identified with the same macroscopic particle is open at this stage. It seems also conceivable that many more states obey the one particle Dirac equation. We postpone this issue to a future investigation. At this point we only emphasize that this feature is present in the standard quantum field theoretic formulation of a Grassmann functional integral for massless Majorana spinors. It is not a particularity of the realization by a classical ensemble for an Ising type model.

\bigskip\noindent
{\bf 4. Observables for excitations of the half-filled

~~vacuum}

A convenient set of classical observables for the characterization of excitations of the half filled vacuum is given by $N'_\gamma(x)$ which takes the value $\big(N'_\gamma(x)\big)_\tau=1/2$ if the bit $(x,\gamma)$ is occupied, and $\big(N'_\gamma(x)\big)_\tau=-1/2$ if it is empty. This classical observable corresponds to the Grassmann operator
\be\label{240A}
\cN'_\gamma \x=\frac12\left[\frac{\partial}{\partial\psi_\gamma\x}
\psi_\gamma\x-\psi_\gamma\x\frac{\partial}{\partial\psi_\gamma\x}\right] 
\ee
for which the basis elements $g_\tau$ are eigenstates
\be\label{XXB}
\cN'_\gamma(x)g_\tau=\big(N'_\gamma(x)\big)_\tau g_\tau,
\ee
with eigenvalues $\pm 1/2$. (The eigenvalue $-1/2$ is found if $g_\tau$ contains a factor $\psi_\gamma(x)$.) We can write 
\be\label{XXD}
\cN'_\gamma(x)=\cN_\gamma(x)-\frac12.
\ee
A local particle number can be defined as 
\be\label{XXE}
\bar\cN'(x)=\sum_\gamma\cN'_\gamma(x)=\sum_\gamma\cN_\gamma(x)-2.
\ee
Its spectrum has the values $(-2,-1,0,1,2)$. We further may employ the total particle number
\be\label{XXF}
\cN'=\sum_x\bar\cN'(x)=\cN-B/2,
\ee
with spectrum $(-B/2,-B/2+1,\dots B/2-1,B/2)$. 

In the complex basis we find the expressions
\ba\label{XXG}
\cN'_1(x)&=&\frac12 \left\{
\left(\frac{\partial}{\partial\zeta_1\x}+\frac{\partial}{\partial\zeta^*_1\x}\right)
\bl\zeta_1(x)+\zeta^*_1\x \br -1\right\}\nn\\
\cN'_2\x&=&\frac12\left\{ \left(
\frac{\partial}{\partial\zeta_1\x}-\frac{\partial}{\partial\zeta^*_1\x}\right)
\bl\zeta_1\x-\zeta^*_1\x\br -1\right\},\nn\\
\cN'_3\x&=&\frac12\left\{\left(\frac{\partial}{\partial\zeta_2\x}+\frac{\partial}{\partial\zeta^*_2\x}\right)
\bl\zeta_2\x+\zeta^*_2\x\br -1\right\},\nn\\
\cN'_4\x&=&\frac12\left\{\left(\frac{\partial}{\partial\zeta_2(x)}-\frac{\partial}{\partial\zeta^*_2\x}\right)
\bl\zeta_2\x-\zeta^*_2\x\br-1\right\},\nn\\
\ea
and we define the combinations
\ba\label{XXH}
\cN'_\uparrow\x&=&\cN'_1\x+\cN'_2\x=\cN_1\x+\cN_2\x-1\nn\\
&=&\frac{\partial}{\partial\zeta_1\x}\zeta_1\x
+\frac{\partial}{\partial\zeta^*_1\x}\zeta^*_1\x-1\nn\\
&=&\frac{\partial}{\partial\psi_1\x}\psi_1\x
+\frac{\partial}{\partial\psi_2\x}\psi_2\x-1,\nn\\
\cN'_\downarrow\x&=&\cN'_3\x+\cN'_4\x=\cN_3\x+\cN_4\x-1\nn\\
&=& \frac{\partial}{\partial\zeta_2\x} \zeta_2\x 
+\frac{\partial}{\partial\zeta^*_2\x}\zeta^*_2\x-1\nn\\
&=&\frac{\partial}{\partial\psi_3\x}\psi_3\x
+\frac{\partial}{\partial\psi_4\x}\psi_4\x-1,
\ea
with 
\be\label{XXI}
\bar\cN'\x=\cN'_\downarrow\x+\cN'_\downarrow\x.
\ee
The half filled vacuum state $g_0$ is an eigenstate of $\cN'_\uparrow\x$ and $\cN'_\downarrow\x$, 
\be\label{XXJ}
\cN'_\uparrow\x g_0=0~,~\cN'_\downarrow\x g_0=0
\ee
and obeys therefore also
\be\label{XXK}
\bar \cN'\x g_0=0~,~\cN' g_0=0.
\ee
On the other hand, $g_0$ is not an eigenstate of the difference observables $\cN'_1\x-\cN'_2\x$ or $\cN'_3\x-\cN'_4\x$. 

The one-particle state \eqref{111F} is an eigenstate of $\cN'$, and similarly for the one-hole state \eqref{111G},
\be\label{XXL}
\cN' g_1=g_1~,~\cN' g_{-1}=- g_{-1}.
\ee
This follows from the commutation relations
\ba\label{XXM}
\big[a^\dagger_\gamma(x),\cN'_\epsilon(y)\big]&=&-a^\dagger_\gamma\x\delta_{\gamma\epsilon}\delta(x,y),\nn\\
\big[a_\gamma\x,\cN'_\epsilon(y)\big]&=&a_\gamma\x\delta_{\gamma\epsilon}\delta(x,y),
\ea
which imply that $a^\dagger_\gamma\x g_0$ and $a_\gamma(x)g_0$ are eigenstates of $\cN'_\uparrow(y)$ and $\cN'_\downarrow(y)$, with eigenvalues $1,0$ or $-1,0$, respectively,
\ba\label{XXN}
\cN'_\uparrow(y)a^\dagger_\gamma\x g_0&=&A_{\uparrow\gamma}\delta(x,y)a^\dagger_\gamma\x g_0,\nn\\
\cN'_\downarrow(y)a^\dagger_\gamma\x g_0&=& A_{\downarrow\gamma}\delta(x,y)a^\dagger_\gamma\x g_0,\nn\\
\cN'_\uparrow(y)a_\gamma\x g_0&=&-A_{\uparrow\gamma}\delta(x,y)a_\gamma\x g_0,\nn\\
\cN'_\downarrow(y)a_\gamma\x g_0&=&-A_{\downarrow\gamma}\delta(x,y)a_\gamma\x g_0,
\ea
where
\ba\label{XX0}
&&A_{\uparrow\gamma}=\left\{ 
\begin{array}{lll}
1&\text{ for }&\gamma=1,2\\
0&\text{ for }&\gamma=3,4
\end{array}\right.,\nn\\
&&A_{\downarrow\gamma}=\left\{
\begin{array}{lll}
0&\text{ for }&\gamma=1,2\\
1&\text{ for }&\gamma=3,4
\end{array}
\right..
\ea
With 
\ba\label{XXP}
\bar \cN'(y)a^\dagger_\gamma \x g_0&=&\delta(x,y)a^\dagger_\gamma\x g_0,\nn\\
\bar \cN'(y)a_\gamma\x g_0&=&-\delta(x,y)a_\gamma\x g_0,
\ea
eq. \eqref{XXL} follows directly.

The Grassmann operators $\bar\cN'\x$ can be used for defining a position operator for a single particle as 
\be\label{XXQ}
\hat X=\int_x x\bar\cN'\x.
\ee
For one-particle states its expectation value is given by 
\be\label{XXR}
\kl \hat X\kr=\int_x xw(x)~,~w(x)=\kl \bar \cN'\x\kr,
\ee
where according to eq. \eqref{G19} one has
\ba\label{XXS}
w\x&=&\int \cD\psi\tilde g_1\bar\cN'\x g_1\\
&=&\int_{y,z} q_\epsilon (y)q_\gamma(z)\int \cD\psi\tilde g_{1,\epsilon}(y)
\bar\cN'\x g_{1,\gamma}(z), \nn
\ea
where
\ba\label{XXT}
g_{1,\gamma}(z)=\sqrt{2}a^\dagger_\gamma(z)g_0
\ea
is normalized according to
\be\label{XXU}
\int\cD\psi \tilde g_{1,\epsilon}(y)g_{1,\gamma}(z)=\delta_{\epsilon\gamma}\delta(y,z).
\ee
Using eq. \eqref{XXP} and the normalization \eqref{XXU} yields
\be\label{XXV}
w(x)=\sum_\gamma q^2_\gamma\x=\sum_\alpha\varphi^*_\alpha\x\varphi_\alpha\x.
\ee
Thus $w(x)\geq 0$ and $\int_x w\x=1$, and we can associate $w\x$ with the probability to find a particle at position $x$.

The probabilistic interpretation of $w\x$ allows us to define operators corresponding to general functions of $x$,
\be\label{XXW}
f(\hat X)=\int_x f\x\bar\cN'\x,
\ee
with expectation value 
\be\label{XXX}
\kl f(\hat X)\kr=\int_x f(x)w(x).
\ee
Since $f(\hat X)$ is a linear combination of operators $\bar \cN'(x)$ and therefore of $\cN_\gamma\x$ it has a direct correspondence with a classical observable which takes in a state $\tau$ the value 
\be\label{XXY}
\bl f(\hat X)\br_\tau=\int_x f\x\big[\sum_\gamma\big (N_\gamma\x\br_\tau-2\big].
\ee
It expectation value obeys the standard statistical law 
\be\label{XXZ}
\kl f(\hat X)\kr=\sum_\tau p_\tau\bl f(\hat X)\br_\tau.
\ee

We define an operator product between operators
\be\label{XX1}
\hat X_l\hat X_{l}=\int_x x_k x_l\bar \cN'\x.
\ee
It maps the classical observables characterized by $(\hat X_k)_\tau$ and $(\hat X_l)_\tau$ to a new observable specified by $(\hat X_k\hat X_l)_\tau$ according to eqs. \eqref{XX1}, \eqref{XXY}. In particular, we may define the dispersion observable
\be\label{XX2}
\Delta^2=\sum_k(\hat X_k-\kl \hat X_k\kr)^2.
\ee
Its expectation value has the familiar probabilistic interpretation 
\be\label{XX3}
\kl\Delta^2\kr=\int_x w\x\sum_k(x_k-\kl x_k\kr)^2.
\ee
It seems rather obvious that the product \eqref{XX1} is appropriate for the description of the distribution of position of a particle. 

The classical observable product reads
\ba\label{XX4}
(\hat X_k\cdot \hat X_l)_\tau &=&(\hat X_k)_\tau (\hat X_l)_\tau\nn\\
&=&\int_{x,y} x_k y_l\bl\bar N'\x\br_\tau\bl\bar N'(y)\br_\tau.
\ea
It coincides with the product \eqref{XX1} only for those classical states $\tau$ for which 
\be\label{XX5}
\bl\bar N'\x\br_\tau\bl\bar N'(y)\br_\tau=\bl\bar N'\x\br_\tau\delta(x,y).
\ee
The classical states $\tau$ for which the probability $p_\tau$ is nonvanishing for a pure one-particle state $g_1$ are precisely of this type. Indeed, the Grassmann basis elements $g_\tau$ which contribute to $g_{1,\gamma}(z)$ in eq. \eqref{XXT} obey $\bar \cN'(z)g_\tau=g_\tau,~\bar \cN'(y\neq z)g_\tau=0$, such that all involved classical states have $\bl\bar N'\x\br_\tau\neq 0$ only for one particular position $x=z$. Extending the definition of the position operator \eqref{XXQ} to more general states the product \eqref{XX1} will no longer coincide with the classical product \eqref{XX4} \cite{3A}.

From eqs. \eqref{XXR}, \eqref{XXV} we infer that the expectation value of the position observable can be computed according to the standard quantum rule from the wave function in position space
\ba\label{XX6}
\kl\hat X_k\kr&=&\int_x\sum_\alpha\varphi^*_\alpha(x)x_k\varphi_\alpha\x,\nn\\
\kl f(\hat X)\kr&=&\int_x\sum_\alpha\varphi^*_\alpha\x f\x\varphi_\alpha\x.
\ea
The product \eqref{XX1} corresponds to the operator product in quantum mechanics and may therefore be called ``quantum product'' \cite{3A}. The formalism of quantum mechanics can be used for a representation of eq. \eqref{XX6} in an arbitrary basis, e.g. the momentum basis where $\hat X$ involves a derivative with respect to momentum.

The definition of the position operator \eqref{XXQ} has been adapted to the one-particle state. For a one-hole state we have to replace $\bar \cN'\x$ by $-\bar\cN'\x$. For this purpose we replace $\bar \cN'\x$ in eq. \eqref{XXQ} by the product $\tilde \cN\x=\bar\cN'\x\cN'$ which coincides with $\bar\cN'\x$ and $-\bar\cN'\x$ for the one-particle or one-hole states, respectively. This extends the definition of the position observable to states which are superpositions of one-particle and one-hole states. We now have
\be\label{XX7}
w\x=\sum_\gamma\bl q^2_\gamma\x+\hat q^2_\gamma\x\br,
\ee
where the correct normalization of $q$ and $\hat q$ corresponds to $\int_xw\x=1$. 

\bigskip\noindent
{\bf 5. Generalized one-particle states}

\medskip
The notion of a position observable is appropriate for generalized one-particle states. (For generalized two-particle states a pair of two positions is appropriate.) For such generalized one-particle states one may replace in eq. \eqref{XXQ} $\bar\cN\x$ or $\tilde \cN\x$ by some operator $J\x$ with the properties 
\be\label{XX7a}
\int_xJ\x=1~,~\kl J\x\kr\geq 0. 
\ee
Then $w\x=\kl J\x\kr$ defines a probability and $f(\hat X)$ can be defined according to eq. \eqref{XXW}. The product \eqref{XX1} will coincide with the classical product only if eq. \eqref{XX5} holds for $J\x$.

There exists a wide class of possible notions of generalized one-particle states and the choice of $J\x$ is not unique. As a first example, which is close to a possible experimental setting, we may cover some region of space by a grid of detectors, with centers placed on a cubic lattice with lattice points $\bar x$ and lattice distance $a$. To any given detector we associate a cube $I_a(\bar x)$ covering all points in the intervals $\bar x_k-a/2<x_k<\bar x_k+a/2$. We define a ``detector observable'' $J_a(\bar x)$ by $\bl J_a(\bar x)\br_\tau=1$ if $\bl\bar N'\x\br_\tau\neq 0$ for any one of the points $x\in I_a\xx$, and $\bl J_a\xx\br_\tau=0$ if $\bl\bar N'\x\br_\tau=0$ for all points in the interval $I_a\xx$. In other words, the detector ``fires'' if a microscopic particle is present at any point of its ``detection region'' $I_a\xx$

A classical state $\tau$ describes a generalized one-particle state if the detector observables obey
\be\label{XX8}
\sum_{\bar x}\bl J_a\xx\br_\tau=1,
\ee
i.e. if precisely one detector of the array fires. We can define the position observable by
\be\label{XX9}
(\hat X_a)_\tau=\sum_{\bar x}\bar x\bl J_a\xx\br_\tau,
\ee
and use the probability that a detector at $\bar x$ fires
\be\label{XX10}
w_a\xx=\kl J_a\xx\kr=\sum_\tau p_\tau\bl J_a\xx\br_\tau,
\ee
where the sum over $\tau$ is restricted to the generalized one particle states. For the associated Grassmann operators ${\cal J}_a\xx$ the basis elements $g_\tau$ are eigenstates with eigenvalues $1$ or $0$ according to 
\be\label{XX11}
{\cal J}_a\xx g_\tau=\bl J_a\xx\br_\tau g_\tau.
\ee

Another example may describe a microscopic particle in a typical situation where many local or ``microscopic'' particles or holes are excited. For this purpose we associate to a given state $\tau$ ``function observables'' $f_{\gamma,\tau}\xx$ \cite{3A}. We choose the value of $f_\gamma$ at $\bar x$ to be proportional to the number of microscopic particles of species $\gamma$ in the interval $I_a\xx$, 
\be\label{XX12}
f_{\gamma,\tau}\xx=c_\tau\sum_{x\in I_a\xx}
\bl N'_\gamma\x\br_\tau,
\ee
with normalization $c_\tau$ chosen such that 
\be\label{XX13}
\sum_{\bar x}\sum_\gamma f^2_{\gamma,\tau}\xx=1.
\ee
We observe that $f_{\gamma,\tau}$ is negative  if more holes than particles are present in the interval $I_a\xx$. For a suitable subset of states $\tau$ that we associate with generalized one particle states we can define 
\be\label{XX14}
\bl J_a\xx\br_\tau=\sum_\gamma f^2_{\gamma,\tau}\xx,
\ee
and use eqs. \eqref{XX9}, \eqref{XX10} for the definition of a position observable. Eq. \eqref{XX5} is no longer obeyed and the ``quantum product'' \eqref{XX1} for observables differs from the classical product \eqref{XX4}. If the grid of points $\bar x$ is fine enough we can associate $f_{\gamma,\tau}\xx$ with functions of a continuous variable $\bar x$. 

The ``mesoscopic wave function'' $f_{\gamma,\tau}\xx$ can be viewed as a set of classical observables, one for every value of $\gamma$ and $\bar x$. Derivatives of $f_{\gamma,\tau}$ with respect to $\bar x$ are therefore also classical observables. In particular, the components of ``mesoscopic momentum''
\be\label{XX15}
(\hat P_k)_\tau =-\sum_{\bar x}\sum_{\gamma,\delta}f_{\gamma,\tau}\xx\tilde I_{\gamma\delta}\partial_k f_{\delta,\tau}\xx
\ee
are classical observables which take fixed values for a given state $\tau$ of the Ising model. We define a mesoscopic density matrix by
\ba\label{XX16}
\rho_{\gamma\delta}(\bar x,\bar y)&=&\kl f_\gamma\xx f_\delta(\bar y)\kr\nn\\
&=&\sum_\tau p_\tau f_{\gamma,\tau}\xx f_{\delta,\tau}(\bar y).
\ea
In terms of $\rho$ the expectation values of position and momentum of mesoscopic particles find an expression familiar from quantum mechanics
\be\label{XX17}
\kl\hat X_k\kr=\text{Tr}(\hat X_k\rho)~,~\kl\hat P_k\kr=\text{Tr}(\hat P_k\rho),
\ee
with trace Tr taken in internal space $(\gamma,\delta)$ and position space $(\bar x,\bar y)$. The operators in eq. \eqref{XX17} are given by
\ba\label{XX18}
(\hat X_k)_{\gamma\delta}(\bar x,\bar y)&=&\delta_{\gamma\delta}\delta(\bar x,\bar y)y_k,\nn\\
(\hat P_k)_{\gamma\delta}(\bar x,\bar y)&=&-\tilde I_{\gamma\delta}\delta(\bar x,\bar y)\frac{\partial}{\partial \bar y_k}.
\ea

A quantum product between the classical observables $(\hat X_k)_\tau$ and $(\hat P_k)_\tau$ is induced \cite{3A} by the product of the associated operators given in eq. \eqref{XX18}. This product is non-commutative. The expectation values of quantum products involving arbitrary powers the mesoscopic observables $\hat X_k$ and $\hat P_k$ can be computed in terms of the density matrix similar to eq. \eqref{XX17}. In contrast, classical products of the mesoscopic observables cannot be computed in terms of $\rho$. We may employ a complex formulation using
\ba\label{XX19}
\varphi^{(a)}_{1,\tau}\xx&=&f_{1,\tau}\xx+if_{2,\tau}\xx,\nn\\
\varphi^{(a)}_{2,\tau}\xx&=&f_{3,\tau}\xx+if_{4,\tau}\xx.
\ea
The complex density matrix
\be\label{XX20}
\rho_{\alpha\beta}(\bar x,\bar y)=\kl \varphi^{(a)}_\alpha\xx\bl\varphi_\beta^{(a)}(\bar x)\br^*\kr
\ee
obeys the standard conditions for a quantum density matrix, i.e. positivity, hermiticity and normalization,
\be\label{XX21}
\rho^\dagger=\rho~,~\text{tr}\rho=1.
\ee
In the complex representation we find the standard representation of position and momentum operators of quantum mechanics (in units with $\hbar=1$)
\be\label{Xx22}
\hat X_k=x_k~,~\hat P_k=-i\frac{\partial}{\partial x_k},
\ee
and therefore the standard commutation relation between $\hat X_k$ and $\hat P_k$. 

The mesoscopic observables $\hat X$ and $\hat P$ and their quantum products are well defined classical observables for the Ising type model and the associated Grassmann functional integral. The question remains if they can be used in practice for a realistic situation of a mesoscopic particle. For example, we may want to describe an atom as a mesoscopic particle, with microscopic scale at the Planck length or even shorter. Indeed, an atom involves many microscopic fluctuations, as gluons in its nucleons etc.. The answer to our question depends to a large extent on the time evolution of the density matrix \eqref{XX16}. This is dictated by the underlying evolution of the probability distribution $\{p_\tau(t)\}$ or the classical wave function $\{q_\tau(t)\}$,
\ba\label{XX23}
\rho_{\gamma\delta}(\bar x,\bar y;t)&=&\sum_\tau\hat f_{\gamma,\tau}\bl\bar x;t\br\hat f_{\delta,\tau}(\bar y;t),\nn\\
\hat f_{\gamma,\tau}\bl\bar x,t\br&=&q_\tau(t)f_{\gamma,\tau}\xx.
\ea
For an isolated mesoscopic particle the evolution equation 
\be\label{XX24}
\partial_t\rho_{\gamma\delta}(\bar x,\bar y)=2\sum_{\tau,\rho}f_{\gamma,\tau}\xx 
f_{\delta,\tau}(\bar y)q_\tau K_{\tau\rho}q_\rho,
\ee
(cf. eq. \eqref{P35}) should be described by a von Neumann equation where the r.h.s. can be expressed in terms of $\rho$. This will certainly not be the case for general wave functions $\{q_\tau\}$. Finding out the characteristic properties of states for which the ``subsystem'' described by $\rho$ is isolated, i.e. for which the time evolution of $\rho$ can  be computed without using information beyond $\rho$, is presumably less a formal task. It rather involves physical understanding as illustrated by our example of atoms. In this context we note that there are many different ways \cite{3A} for the definition of function observables $f_{\gamma,\tau}\xx$ which obey the normalization condition \eqref{XX13} and therefore define a density matrix \eqref{XX16}. 

\bigskip\noindent
{\bf 6. Particle-hole identification}

\medskip
The wave function for single particle or a single hole obey the same Dirac equation \eqref{H13}. Similarly, multi-particle states and multi-hole states obey the same evolution equation. For a particle-hole symmetric vacuum it is therefore easy to describe particle-hole symmetric excitations. The eigenstates of the ${\cal C}_{ph}$-operator are given for the one-particle and one-hole states by 
\be\label{XX25}
q_{\pm,\gamma}\x=\frac{1}{\sqrt{2}}\bl q_\gamma \x\pm\hat q_\gamma\x\br,
\ee
and similarly for multi-particle states. The eigenstates of ${\cal C}_{ph}$  with eigenvalue $+1$ have the property that the time evolution of the Grassmann wave function $g(t)$ is smooth for neighboring time points $t$ and $t+\epsilon$, if $g(t)$ contains only elements with an even number of Grassmann variables $\psi_2$ or $\psi_4$. For an odd number it changes sign. This follows from eq. \eqref{188A} which implies ${\cal C}g_\tau=\tilde \cK g_\tau$ in eq. \eqref{P15}. A particle-hole symmetric initial state ${\cal C}_{ph}g_{in}=g_{in}$ remains particle-hole symmetric for all even times, ${\cal C}_{ph}g(t)=g(t)$.

Particle-hole symmetric states are linear superpositions of particle and hole wave functions which do not allow anymore any distinction between particles and holes. A ''one-particle state'' should therefore be seen as a particle-hole superposition. It can be simply obtained by identifying $\hat q_\gamma\x=q_\gamma\x$, with an appropriate normalization. In terms of the classical statistical ensemble for the Ising type model this simply means that the probabilities for the particle and hole excitations of the particle hole symmetric vacuum state are equal.

\section{Conclusions and discussion}
\label{Conclusionsanddiscussion}
In conclusion, the real Grassmann functional integral based on the action \eqref{F1}, or its regularized discretized version \eqref{N1}, \eqref{C.1}, \eqref{C.4}, defines a fully consistent quantum field theory for free massless Majorana or Weyl spinors. No complex numbers are needed at this stage. Nevertheless, the time evolution of the associated multi-particle wave function is unitary. One-particle wave functions obey the Dirac equation. Multi-particle wave functions are totally antisymmetric, as appropriate for fermions.

The central result of this paper states that this system can equivalently be described by a classical statistical ensemble for an Ising type model, with a suitable time evolution law for the probability distribution. This constitutes an explicit realization of quantum mechanics by a classical statistical ensemble, in contrast to widespread belief that this is impossible. When formulated in terms of the probabilities the time evolution law \eqref{G14} is non-linear. It takes, however, a very simple linear form once formulated in terms of the classical wave function which corresponds to a ``root'' of the probability distribution.

We explicitly indicate the probability distribution for the Ising model which corresponds to a quantum state for a propagating fermion. A particularly simple example is a probability distribution that vanishes whenever more than one bit is occupied or if all bits are empty. The states $\tau$ of the Ising model for which $p_\tau$ differs from zero can then be associated with the single occupied bits $(x,\gamma)$, $p_\tau\widehat{=}p(x,\gamma)$. The normalization of the corresponding probabilities $p(x,\gamma)$ obeys $\sum_x\sum_\gamma p(x,\gamma)=1$. A state with a fixed momentum $p_3>0$ in the $3$-direction is given by a probability distribution
\ba\label{224}
p_{(0,0,p_3)}(x,1)&=&L^{-3}\cos^2\{p_3(t+x_3)\},\nn\\
p_{(0,0,p_3)}(x,2)&=&L^{-3}\sin^2\{p_3(t+x_3)\},
\ea
and $p(x,3)=p(x,4)=0$. The probability oscillates in time and space between species $1$ and $2$. For states with two occupied bits one finds the interference effects characteristic for fermions.

The evolution of the probability distribution \eqref{224} can be expressed in a simple way by a wave function
\ba\label{225}
q(x,1)&=&L^{-3/2}\cos \{p_3(t+x_3)\},\nn\\
q(x,2)&=&L^{-3/2}\sin \{p_3(t+x_3)\},\nn\\
p(x,\gamma)&=&q^2(x,\gamma).
\ea
(There is a freedom of choice for the overall signs of $q(x,1)$ and $q(x,2)$, while continuity and differentiability of $q$ fix the relative signs for different $x$ and $t$.) The complex structure is introduced by a complex wave function
\ba\label{241A}
\varphi(x)&=&q(x,1)+iq(x,2)\nn\\
&=&L^{-3/2}\exp \{ip_3(t+x_3)\},
\ea
with
\be\label{226}
\sum_x\varphi^*(x)\varphi(x)=1.
\ee
It obeys the simple evolution law for a relativistic particle propagating in the $x_3$-direction
\be\label{227}
\partial_t\varphi=\partial_3\varphi.
\ee
Our setting generalizes this evolution equation for arbitrary momentum and spin, as well as for multi-particle states, in a Lorentz invariant way. It describes free massless Weyl or Majorana fermions. The complex structure of quantum field theory can be associated in a natural way to the equivalence of Majorana and Weyl spinors. 

Our formulation shows the particle-wave duality characteristic for quantum mechanics. The particle aspects correspond to the discrete spectrum of observables. For example, the observable $N_\gamma(x)$ counts for a given state of the classical ensemble if the bit $(x,\gamma)$ is occupied or empty. It can only take the values $N_\gamma(x)=1,0$ - either a particle of species $\gamma$ is present at $x$ or not. According to the standard rules of classical statistics each measurement of $N_\gamma(x)$ can only yield the values one or zero. In other words, in any given state $\tau$ this observable has a fixed value $\big(N_\gamma(x)\big)_\tau=1,0$. On the other side, the probability distribution is typically a continuous function if we take the continuum limit for time and space. The wave aspects are related to the probabilities. The characteristic interference phenomena arise for an evolution law that is linear in $q$. 

In our approach the axioms for the relation between the formalism of quantum mechanics and the outcome of measurements follow directly from the standard axioms of probability theory or classical statistics. All measurements should be formulated as measurements of classical observables in a classical statistical ensemble. We have shown that this can indeed be done for all observables that can be constructed as functions of $N_\gamma(x)$. The expectation values of all such observables only depend on the probabilities $p_\tau$, while the signs $s_\tau$ of the wave function, $q_\tau=s_\tau \sqrt{p_\tau}$, are irrelevant. 

In the continuum limit, functions of the position observables can easily be constructed from $N_\gamma\x$. For Ising states with only one occupied bit one uses the position observable
\be\label{CA}
\vec X=\sum_{x,\gamma}\vec x N_\gamma(\vec x).
\ee
This can be generalized to one-particle excitations of a more general vacuum, as the half filled vacuum - in this case one may add a constant to $N_\gamma(\vec x)$ in order to guarantee $\sum_\gamma N_\gamma \x=0$ for the vacuum state. The observables $f(\vec X)$ replace in eq. \eqref{CA} $\vec x$ by $f(\vec x)$. Expectation values of $f(\vec X)$ at arbitrary $t$ can be equivalently computed from the classical ensemble for the generalized Ising model or from the quantum wave function. The relation between the expectation values and the results of measurements can be directly inferred from the standard classical statistical setting. In our setting the interference effects characteristic for quantum mechanics do not pose any conceptual challenge. For example, the interference pattern of a double slit experiment simply reflects the probabilities to find a particle at a certain position of the detector. In principle, these probabilities can be computed using only the evolution law for the probability distribution. In practice, the concept of a ``classical wave function'' $\{q_\tau(t)\}$, with $\{p_\tau(t)\}=\{q^2_\tau(t)\}$, is very convenient, however.

For sequences of measurements one needs to define the appropriate ``measurement correlation'' which is based on conditional probabilities \cite{GR}. There are may different consistent possible choices of measurement correlations or the associated observable products. In principle, the choice of the correct correlation function depends on the specific setting for the measurements, reflecting how a first measurement affects the outcome of a second one. For many situations the measurement correlation differs from the classical correlation function. We have explicitly indicated an example of a mesoscopic particle where momentum and position are both classical statistical observables of the Ising model which can be expressed in terms of $N_\gamma\x$. For this example we have defined a non-commuting product of observables such that the associated correlation functions reflect precisely the quantum correlations.

We emphasize that all formal and mathematical aspects of our discussion of the classical statistical ensemble for the generalized Ising model are specified by the regularized quantum field theory for Majorana spinors. Inversely, none of our formal findings depends on the map to the classical statistical ensemble. They are properties of a standard Grassmann functional integral describing a quantum field theory for fermions. This concerns, in particular, the issue of different vacua raised in sect. \ref{Particle-holeconjugation}. Our setting can be generalized to Dirac particles in an external electromagnetic field which will be described in a separate publication. In this case the single particle wave function will evolve in presence of a potential according to the Dirac-equation. It remains to be seen if the mapping to an equivalent Ising type classical statistical ensemble can also be used for practical purposes beyond the conceptual advances in the understanding of quantum physics.

\section*{APPENDIX A: DISCRETIZATION}
\renewcommand{\theequation}{A.\arabic{equation}}
\setcounter{equation}{0}

{\bf 1. Lattice Fourier transform}

\medskip
The lattice Fourier transform expands (we omit the index $\gamma$)
\be\label{C.10}
\psi_m=\frac{1}{\sqrt{2L^3}}
\sum_q\big\{\cos(\Delta\vec m\vec q)\psi_R(\vec q)-\sin
(\Delta\vec m\vec q)\psi_I(\vec q)\big\}
\ee
with Fourier modes $\psi_{R,I}(\vec q)$ constrained by
\be\label{C.11}
\psi_R(-\vec q)=\psi_R(\vec q)~,~\psi_I(-\vec q)=-\psi_I(\vec q).
\ee
The number of different $j_k$ is $L$, such that the number of Grassmann variables $\psi_R(q),\psi_I(q)$ for each species remains $L^3$, since the doubling into $\psi_R$ and $\psi_I$ is compensated by the identifications for $\psi(q)$ and $\psi(-q)$ in eq. \eqref{C.11}. The transformation \eqref{C.10} can therefore be regarded as a similarity transformation among the $4L^3$ Grassmann variables. We note that $q_k=0$ or $q_k=\pi/\Delta$ is not contained among the possible values in eq. \eqref{C.10}. The minimal and maximal values of $q_k$ are $-\pi/\Delta+\pi/l$ and $\pi/\Delta-\pi/l$. In the continuum limit this amounts to $|q_k|\leq\pi/\Delta$. We may consider $q_k$ as periodic variables by identifying $q_k$ and $q_k+2\pi/\Delta$. The mode with $\vec j=(0,0,0)$ corresponds to
\ba\label{C.13}
\psi(x)&=&\frac{1}{\sqrt{2L^3}}\left\{\cos\left(
\frac{\pi(x_1+x_2+x_3)}{l}\right)\psi_R\right.\nn\\
&&\left.-\sin\left(\frac{\pi(x_1+x_2+x_3)}{l}\right)\psi_I\right\}
\ea
and is antiperiodic in all coordinates $x_k$. 

We may employ complex variables $\psi(q)=\frac{1}{\sqrt{2}}\big\{\psi_R(q)+i\psi_I(q)\big\}$, with a standard complex discrete Fourier transform 
\ba\label{C.14}
\psi_m=L^{-\frac32}\sum_q e^{i\Delta\vec m\vec q}
\psi(q)~,~\psi(-q)=\psi^*(q).
\ea
The identities (for $m,k,N$) integer)
\ba\label{B6}
&&\sum^N_{m=-N}\exp\left\{
\frac{2\pi ikm}{2N+1}\right\}=
(2N+1)\delta_{k,0~mod~2N+1},\nn\\
&&\sum^{N-1}_{m=-N}\exp\left\{
\frac{\pi i km}{N}\right\}=2N\delta_{k,0~mod~ 2N},
\ea
imply the relations
\ba\label{B24}
&&\sum_m\exp\big(i\Delta\vec m(\vec q-\vec p)\big)=L^3\bar\delta_{q,p},\nn\\
&&\sum_m\cos (\Delta\vec m\vec q)\cos(\Delta\vec m\vec p)=
\frac{L^3}{2}(\bar \delta_{q+p,0}+\bar\delta_{q,p}),\nn\\
&&\sum_m\sin(\Delta\vec m\vec q)\sin(\Delta\vec m\vec p)=\frac{L^3}{2}
(-\bar\delta_{q+p,0}+\bar\delta_{q,p}),\nn\\
&&\sum_m\cos(\Delta\vec m\vec q)\sin(\Delta\vec m\vec p)=0,
\ea
with $\bar\delta_{q,p}=1$ for $j_{q,k}=j_{p,k}$ mod $L$, and zero otherwise. The identities \eqref{B6}, \eqref{B24} allow us to establish
\ba\label{C.17}
&&\sum_x\varphi(x)\psi(x)=\sum_m\varphi_m\psi_m\nn\\
&&\quad =\frac12\sum_q\big\{\varphi_R(q)\psi_R(q)+\varphi_I(q)\psi_I(q)\big\}\\
&&\quad =\sum_q\varphi^*(q)\psi(q)=\sum_{q_{>}}\varphi_R(q)
\psi_R(q)+\varphi_I(q)\psi_I(q)\big\}.\nn
\ea
In the last line we employ $q_>$ which may be defined by the condition $q_3>0$. The linear transformation between the independent spinors $\psi_{R,\gamma}(q_{>}),\psi_{I,\gamma}(q_>)$ and $\psi_\gamma(x)$ is therefore a rotation with unit Jacobian, such that the functional measure in terms of $\psi_{R,I}(q_>)$ and $\psi(x)$ is the same. (This may be seen by taking $\psi=\psi(t+\epsilon)$ and $\varphi=\psi(t)$.)

\bigskip\noindent
{\bf  2. Next neighbor interactions}

\medskip
For neighboring lattice sites, with $\Delta_k$ the unit lattice vector in the $k$ direction, $|\Delta_k|=\Delta$, one obtains the relation
\ba\label{C.18}
&&\sum_x\varphi(x)\psi(x+\Delta_k)\nn\\
&&\qquad=\frac12\sum_q\Big\{\cos(\Delta q_k)\big[\varphi_R(q)\psi_R(q)
+\varphi_I(q)\psi_I(q)\big]\nn\\
&&\qquad-\sin(\Delta q_k)
\big[\varphi_R(q)\psi_I(q)-\varphi_I(q)\psi_R(q)\big]\Big\}\nn\\
&&\qquad=\sum_q  e^{i\Delta q_k}
\varphi^*(q)\psi(q).
\ea
This can be used in order to show that the derivative \eqref{C.6}
obeys
\ba\label{C.20}
&&\sum_x\varphi(x)\tilde\partial_k\psi(x)\\
&&\qquad=-\frac{1}{2\Delta}
\sum_q\sin(\Delta q_k)\big[\varphi_R(q)\psi_I(q)-\varphi_I(q)\psi_R(q)\big].\nn
\ea
We may also define
\be\label{C.21}
\tilde\delta_k\psi(x)=\frac{1}{2\Delta}
\big\{\psi(x+\Delta_k)+\psi(x-\Delta_k)-2\psi(x)\big\},
\ee
such that 
\ba\label{C.22}
&&\sum_x\varphi(x)\tilde \delta_k\psi(x)\\
&&=\frac{1}{2\Delta}\sum_q\big[\cos(\Delta q_k)-1\big]
\big[\varphi_R(q)\psi_R(q)+\varphi_I(q)\psi_I(q)\big].\nn
\ea

Our version of a lattice derivative will be introduced by considering the combination of Grassmann variables $\tilde F_\gamma(x)$ similar to eq. \eqref{C.4},
\be\label{C.23}
\tilde F_\gamma(x)=\sum_k\big\{(T_k)_{\gamma\delta}\tilde\partial_k\psi_\delta(x)
+aS_{\gamma\delta}\tilde\delta_k\psi_\delta\langle x\rangle\big\}.
\ee
This can be expressed as a linear transformation acting on $\psi(x)$
\be\label{C.24}
\tilde F_\gamma\big[\psi(t+\epsilon);x\big]=\sum_y
P_{\gamma\delta}(x,y)\psi_\delta(t+\epsilon,y),
\ee
with
\ba\label{C.25}
P_{\gamma\delta}(x,y)&=&\frac{1}{2\Delta}\sum_k
\Big\{(T_k)_{\gamma\delta}\big[\delta(x,y-\Delta_k)\nn\\
&-&\delta(x,y+\Delta_k)\big]\\
&&\hspace{-2.1cm}
+aS_{\gamma\delta}\big[\delta(x,y-\Delta_k)+\delta(x,y+\Delta_k)-2\delta(x,y)
\big]\Big\}.\nn
\ea
We want $P$ to be an antisymmetric matrix
\be\label{C.26}
P_{\delta\gamma}(y,x)=-P_{\gamma\delta}(x,y).
\ee
This is achieved for an antisymmetric matrix $S$, i.e. $S_{\delta\gamma}=-S_{\gamma\delta}$. We will further require the properties
\be\label{C.27}
S^2=-1~,~\{S,T_k\}=0~,~S^T=-S. 
\ee
This is obeyed for $S=\pm\gamma^0$ or $S=\pm\tilde I\gamma^0$.

\bigskip\noindent
{\bf  3. Lattice dispersion relation}

\medskip
The combination $\tilde F_\gamma$ in eq. \eqref{C.23} is chosen such that ``lattice doublers'' are avoided. Our setting can therefore describe a single species of Majorana or Weyl fermions. With
\ba\label{C.28}
&&\sum_x\psi_\gamma(t,x)\tilde F_\gamma\big[\psi(t+\epsilon);x\big]=-
\frac{1}{2\Delta}\sum_q\sum_k\nn\\
&&\quad\Big\{\sin(\Delta q_k)\big[\psi_{R,\gamma}(t,q)(T_k)_{\gamma\delta}\psi_{I,\delta}
(t+\epsilon, q)\nn\\
&&\qquad-\psi_{I,\gamma}(t,q)(T_k)_{\gamma\delta}\psi_{R,\delta}
(t+\epsilon,q)\big]\nn\\
&&\quad+a\big[1-\cos(\Delta q_k)\big]
\big[\psi_{R,\gamma}(t,q)S_{\gamma\delta}\psi_{R,\delta}
(t+\epsilon,q)\nn\\
&&\qquad+\psi_{I,\gamma}(t,q)S_{\gamma\delta}\psi_{I,\delta}(t+\epsilon,q)
\big]\Big\},
\ea
we can express the Fourier transform of $P$ as a matrix in the space of $\big(\psi_R(q),\psi_I(q)\big)$, which is diagonal in momentum space,
\be\label{C.29}
\sum_x\psi_\gamma(t,x)\tilde F_\gamma\big[\psi(t+\epsilon);x\big]=\frac12
\sum_q\tilde \psi^T_\gamma(t,q)P_{\gamma\delta}(q)\tilde\psi_\delta(t+\epsilon,q).
\ee
Here we use
\be\label{C.30}
\tilde \psi_\gamma(q)=\left(\begin{array}{c}
\psi_{R,\gamma}(q)\\\psi_{I,\gamma}(q)\end{array}\right),
\ee
and $P(q)=-P^T(q)$ obeys
\ba\label{C.31}
P(q)=-\frac{1}{\Delta}\sum_k
\left(\begin{array}{lll}
a(1-c_k)S&,&T_k s_k\\-T_ks_k&,&a(1-c_k)S\end{array}\right),
\ea
with
\be\label{C.32}
s_k=\sin(\Delta q_k)~,~c_k=\cos(\Delta q_k).
\ee

The properties \eqref{C.27} of the matrix $S$ ensure that $P^2$ is proportional to the unit matrix,
\be\label{C.33}
P^2(q)=-\frac{1}{\Delta^2}
\Big\{\sum_k s^2_k(q)+a^2\big(\sum_k\big[1-c_k(q)\big]\big)^2\Big\}.
\ee
All eigenvalues of $P^2$ are thus negative or zero. For $a\neq 0$ zero eigenvalues only occur if for all $k$
\be\label{C.34}
s_k(q)=0~,~c_k(q)=1.
\ee
This is realized only for $\vec q=0$. This contrasts to the case $a=0$ where the zeros of $P(q)$ for both $q_k=0$ and $q_k=\pi/\Delta$ lead to the familiar doubling problem for lattice fermions. No doubling is present for $a\neq 0$, since 
\be\label{C.35}
P^2\left(\frac\pi\Delta,0,0\right)=-\frac{4a^2}{\Delta^2},
\ee
and similar for other modes where $q_k=\pi/\Delta$. Except for $\vec q=0$, where $P(\vec q)=0$, one finds that all eigenvalues of $P(q)$ are purely imaginary. 

The matrix $P(q)$ is closely related to the inverse fermion propagator in momentum space and to the dispersion relation which relates frequencies or energies $\omega$ to $\vec q$ (cf. main text). The energy eigenvalues obey
\be\label{C.36}
\big (\omega^2+P^2(\vec q)\big)\tilde \psi(q)=0
\ee
or a dispersion relation
\be\label{C.37}
\omega^2(\vec q)=-P^2(\vec q).
\ee
All squared frequencies $\omega^2$ are strictly positive except for $\vec q=0$ where $\omega(\vec q=0)$. For small $|\vec q|$ one finds the relativistic dispersion relation $|\omega|=|\vec q|$. 

\bigskip\noindent
{\bf  4. Lattice derivative and continuum limit}

\medskip
Since $\tilde F_\gamma(x)$ as defined by eq. \eqref{C.23} has all the desired properties we define the lattice derivative $\partial_k$ by
\be\label{C.38}
\tilde F_\gamma(x)=\sum_k(T_k)_{\gamma\delta}\partial_k\psi_\delta(x),
\ee
and identify
\be\label{C.39}
\partial_k=\tilde\partial_k-\frac a2\sum_{l,m}\epsilon_{klm}T_lT_m\tilde I S\tilde \delta_k.
\ee
For definiteness we may take $a=1/2$. We observe that this lattice derivative is compatible with the cubic lattice symmetry of the space lattice. The object $\sum_k\tilde\delta_k$ transforms proportional to a lattice Laplacian and $S, \tilde I$ are invariant, such that $\tilde F_\gamma(x)$ has the same transformation property as $\psi_\gamma(x)$. 

The continuum limit is associated to the behavior of $P(q)$ close to its zeros. It can be taken as the limit $\Delta\to 0$ for a fixed value of $\vec q$, and we concentrate on momenta where $P^2(q)$ \eqref{C.33} remains finite in this limit. This happens for $\vec q^2\ll \Delta^{-2}$, where in eq. \eqref{C.33} we can replace $s_k=\Delta q_k~,~1-c_k=\Delta^2 q^2_k/2$. The diagonal elements of $P(q)$ vanish in the continuum limit and we observe the linear dispersion relation
\be\label{C.40}
P(q)=\sum_k\left(\begin{array}{cc}
0,&-T_kq_k\\T_k q_k,&0\end{array}\right)~,~
P^2(q)=-\vec q^2.
\ee
As we discuss in sect. \ref{Particlestates} the continuum time evolution equation for a one particle wave function $q(x)$ is given by the Dirac equation
\be\label{C.41}
\partial_tq(x)=\sum_kT_k\partial_k q(x)=\int_y P(x,y)q(y).
\ee
In momentum  space, with $\tilde q(p)=\big[q_R(p),q_I(p)\big]$, this reads
\be\label{C.42}
\partial_t\tilde q(p)=P(p)\tilde q(p).
\ee
In a complex formulation with $q_c(p)=\frac{1}{\sqrt{2}}\big(q_R(p)+iq_I(p)\big)$ eq. \eqref{C.42} yields
\be\label{C.43}
\partial_t q_c(p)=i\sum_k p_k T_k q_c(p).
\ee

\bigskip\noindent
{\bf  5. Symmetries}

\medskip
We emphasize that our definition of the lattice derivative \eqref{C.39} or the corresponding definition \eqref{C.23} for $\tilde F_\gamma$ is not compatible with the complex structure \eqref{186B}. No matrix $S$ exists which anticommutes with all $T_k$ and commutes with $\tilde I$. (Otherwise we could find a complex $2\times 2$ matrix which anticommutes with all three Pauli matrices $\tau_k$, which is not possible.) We will take for definiteness
\be\label{C.44}
S=\gamma^0~,~\{S,\tilde I\}=0,
\ee
which makes the discrete formulation compatible with the parity transformation \eqref{85Ba}  and time reversal \eqref{166A}. The action of the lattice derivative \eqref{C.39} on the complex Grassmann variable $\zeta$ becomes then
\be\label{C.45}
\partial_k\zeta=\tilde \partial_k\zeta-\frac {ia}{2}\sum_{l,m}\epsilon_{klm}\tau^*_l\tau^*_m\tau_2\tilde \delta_k\zeta^*.
\ee
For $a\neq 0$ this mixes $\zeta$ and $\zeta^*$.

This incompatibility of $\partial_k$ with the complex structure is closely related to the observation that the ``regulator term'' $\sim a$ violates the symmetry of continuous rotations \eqref{71A}, \eqref{71B}. Indeed, the transformation \eqref{71B} implies
\ba\label{C.46}
&&\delta\big(\psi_\gamma F_\gamma(\psi)\big)=\alpha
\big[(\tilde I\psi)_\gamma F_\gamma(\psi)+\psi_\gamma F_\gamma(\tilde I\psi)\big]\nn\\
&&=\alpha\psi_\gamma\big[-\big(\tilde IF(\psi)\big)_\gamma+F_\gamma(\tilde I\psi)\big].
\ea

From $\{S,\tilde I\}=0$ one infers 
\ba\label{C.47}
\tilde I\tilde F&=&\tilde I\sum_k(T_k\tilde\partial_k+aS\tilde \delta_k)\psi\nn\\
&=&\sum_k(T_k\tilde\partial_k-aS\tilde\delta_k)\tilde I\psi\neq 
\tilde F[\tilde I\psi].
\ea
In our case of free fermions this symmetry violation can be neglected in the continuum limit. 

The violation of the $U(1)$-symmetry acting on $\zeta$ by the regulator term reflects the well known fact that the $U(1)$ symmetry for a single Weyl spinor may remain anomalous for an extended theory even in the continuum limit. Our functional measure is invariant under the transformation \eqref{71A}, \eqref{71B}. If one could find a regulated action for a single Weyl spinor that is invariant under $U(1)$ transformations this would constitute an anomaly free realization of this symmetry. In our setting, the $U(1)$-symmetry is preserved only for $a=0$. In this case, however, we encounter fermion doubling and the model no longer describes a single Weyl spinor. This is consistent with the expectations from the Nielson-Ninomiya theorem \cite{NN}.

\bigskip\noindent
{\bf  6. Rotation of Grassmann variables}

\medskip
The quantity $F_\gamma(x)$ in eq. \eqref{C.4} coincides with $\tilde F_\gamma(x)$, as defined by eq. \eqref{C.23}, only in the limit $\epsilon\to 0$. For finite $\epsilon$ we use the relation \eqref{C.24} in a generalized matrix notation
\be\label{C.48}
\tilde F=P\psi
\ee
in order to define
\be\label{C.49}
F=\frac 1\epsilon\big[1-\exp(-\epsilon P)\big]\psi
=P\psi+0(\epsilon^2).
\ee
(This yields a corresponding modification of the lattice derivative $\partial_k$ for finite $\epsilon/\Delta$.) The reason for this modification is that we can now write
\be\label{C.50}
B=\psi-\epsilon F=\exp (-\epsilon P)\psi=R\psi.
\ee
Since $P^T=-P$ we infer that $R$ is indeed a rotation, $R^TR=1$.

\section*{APPENDIX B: FUNCTIONAL INTEGRAL WITH CONJUGATE GRASSMANN VARIABLES}
\renewcommand{\theequation}{B.\arabic{equation}}
\setcounter{equation}{0}

In this appendix we reformulate the Grassmann functional integral \eqref{N9} with action \eqref{N1}-\eqref{C.8} on a coarse grained time lattice with time points $t_m,m\in {\mathbbm Z},t_{m+1}-t_m=2\epsilon$. In sect. \ref{Timeevolution} we have distinguished between even and odd time points. The coarse grained lattice only retains the even time points. The Grassmann variables for odd time points will be associated with a different set of Grassmann variables at even $t$ by defining for $t$ even
\be\label{A.1}
\psi_\gamma(t+\epsilon,x)=\hat\psi_\gamma(t,x).
\ee
Since all bilinears in eq. \eqref{A17} involve one Grassmann variable at even $t$ and another one at odd $t$, the action on the coarse grained lattice will involve terms with one factor $\psi$ and one factor $\hat\psi$. This formulation will be close to the description of classical probability distributions by Grassmann functional integrals developed in ref. \cite{CWF}, with $\hat\psi$ the conjugate Grassmann variable of $\psi$. Certain properties, as symmetry transformations between $\psi$ and $\hat\psi$, are better visible in this formulation. 

\bigskip\noindent
{\bf   1. Conjugate spinors}

\medskip
We start with the ``trivial theory'' by setting $F_\gamma=0$ in eq. \eqref{C.8}. The action can now be written in the form
\ba\label{A.2}
S&=&\sum^{t_f-2\epsilon}_{t=t_{in}}L(t),\nn\\
L(t)&=&\sum_x\big\{\psi_\gamma(t,x)\psi_\gamma(t+\epsilon,x)\nn\\
&&+\psi_\gamma(t+\epsilon,x)\psi_\gamma(t+2\epsilon,x)\big\}\nn\\
&=&\sum_x\hat\psi_\gamma(t,x)
\big[\psi_\gamma(t+\tilde\epsilon,x)-\psi_\gamma(t,x)\big],
\ea
with $\tilde \epsilon=2\epsilon$ and $\sum_t=\sum_m=\tilde\epsilon^{-1}\int_t$. This yields the dynamical term of the action in the formalism of ref. \cite{CWF}. For $\tilde \epsilon\to 0$ the continuum version reads $S=\int_{t,x}\hat\psi_\gamma\partial_t\psi_\gamma$. Here we have chosen $t_f$ even. We next add the term involving $F_\gamma$, namely
\ba\label{A.3}
&&\Delta L(t)=-\epsilon\sum_x\big\{\psi_\gamma(t,x)
(T_k)_{\gamma\delta}\partial_k\hat\psi_\delta(t,x)\nn\\
&&+\hat\psi_\gamma(t,x)(T_k)_{\gamma\delta}
\partial_k\psi_\delta(t+2\epsilon,x)\big\}\nn\\
&&+(A-1)\sum_x\big\{\psi_\gamma(t,x)\hat\psi_\gamma(t,x)+
\hat\psi_\gamma(t,x)\psi_\gamma(t+2\epsilon,x)\big\}\nn\\
&&=-\tilde \epsilon\sum_x\hat\psi_\gamma(t,x)(T_k)_{\gamma\delta}\partial_k\psi_\delta
(t+\tilde\epsilon,x)\nn\\
&&+(\tilde A-1)\sum_x\hat\psi_\gamma(t,x)\psi_\gamma(t+\tilde\epsilon,x)+C(t),
\ea
with $\tilde A^2=1-3\tilde\epsilon^2/(2\Delta^2)$. The correction term $C(t)$ vanishes in the continuum limit $\tilde\epsilon\to 0$, where
\be\label{F1a}
S=\int_{t,x}\big\{\hat\psi_\gamma\partial_t\psi_\gamma-\hat\psi_\gamma
(T_k)_{\gamma\delta}
\partial_k\psi_\delta\big\}.
\ee
The conjugate Grassmann variables $\hat\psi_\gamma$ obey the same Lorentz-transformations as $\psi_\gamma$.
and the manifestly Lorentz invariant form of $S$ can be written as
\be\label{F8a}
S=-\int_{t,x}\bar\psi\gamma^\mu \partial_\mu \psi~,~\bar\psi=\hat\psi^T\gamma^0.
\ee
We observe the Lorentz invariance of $\int_x\hat\psi_\gamma(x)\psi_\gamma(x)$. Together with the invariance of the functional measure $\int \cD\psi\cD\hat\psi$ this guarantees that also the measure $\int{D}\psi=\int {D}\psi{\cal D}\hat\psi\exp 
\big\{\int_x\hat\psi_\gamma(x)\psi_\gamma(x)\big\}$ is invariant. Therefore the relations between $\hat\psi_\gamma$ and $\psi_\gamma$ are the same in all systems related by Lorentz transformations. 

\bigskip\noindent
{\bf   2. Wave function}

\medskip
On the coarse grained time lattice (with only even $t$) our formulation has the structure of a Grassmann functional integral with conjugate Grassmann variables, which has been discussed extensively in ref. \cite{CWF}. The construction of the classical wave function and the classical probability density can be taken over. We will omit the correction term $C(t)$ in eq. \eqref{A.3}. (This corresponds to a different regularization of the functional integral.) The results in the continuum limit should not depend on this. We also rename $\tilde\epsilon\to \epsilon$, such that the regularized action reads

\ba\label{G1}
S&=&\sum^{t_f}_{t'=t_{in}}L(t'),\nn\\
L(t')&=&\int_x\Big\{\hat\psi(t',x)\big[\psi(t'+\epsilon,x)-\psi(t',x)\big]\nn\\
&&-\epsilon\hat\psi(t',x)T_k\partial_k\psi(t'+\epsilon,x)\Big\}.
\ea
(We omit to write the spinor indices $\gamma$.) As before, we also may place $x$ on a cubic lattice with lattice distance $\Delta$ and use the lattice derivative \eqref{C.5}. In eq. \eqref{G1} we have extended the summation to include $t'=t_f$, with
\be\label{AA1}
L(t_f)=-\sum_x\hat\psi(t_f,x)\psi(t_f,x).
\ee

In consequence, the partition function is now defined by the functional integral
\be\label{G3}
Z=\int {\cal D}\psi{\cal D}\hat\psi\hat g_f\big[\hat\psi(t_f)\big]e^{-S}
g_{in}\big[\psi(t_{in})\big].
\ee
Here the functional integral $\int \cD\psi\cD\hat\psi$ is an integration over all Grassmann variables $\psi$ and $\hat\psi$ (for all $t_{in} \leq t'\leq t_f)$. The particular probability distribution corresponding to this functional integral will be specified by the Grassmann element $g_{in}\big[\psi_\gamma(t_{in},x)\big]$, which can be considered as a boundary term. The element $\hat g_f\big[\hat \psi_\gamma(t_f,x)\big]$ is related to $\bar g\big[\psi(t_f)\big]$ by the additional integration over $\hat\psi(t_f)$ - not present in eq. \eqref{N9} - together with the additional term \eqref{AA1},
\ba\label{AA2}
\int\cD\hat\psi(t_f)\hat g_f\big[\hat\psi(t_f)\big]\exp
\big\{\sum_x\hat\psi(t_f,x)\psi(t_f,x)\big\}=\bar g_f.\nn\\
\ea
We have seen in sect. \ref{Timeevolution}, eq. \eqref{P32a} that $\bar g_f$ is conjugate to $g\big[\psi(t_f)\big]$. In turn $\hat g_f\big[\hat\psi(t_f)\big]$ is conjugate to $g\big[\psi(t_f)\big]$ in the formulation with conjugate spinors \cite{CWF}. As a consequence, it obtains from $g_f\big[\psi_\gamma(t_f,x)\big]$ by replacing each Grassmann variable $\psi_\gamma(t_f,x)\to \hat\psi_\gamma(t_f,x)$, and by a total reordering of all Grassmann variables. In turn, $g_f$ will be related to $g_{in}$ as a solution of a time evolution equation and is therefore not an independent quantity. 

For some given time $t$ we split
\ba\label{G2}
S&=&S_<+S_>-\int_x\hat\psi(t,x)\psi(t,x),\nn\\
S_<&=&\sum_{t_{in}\leq t'<t}L(t'),\\
S_>&=&\sum_{t\leq t'<t_f}
L(t')+\int_x\hat\psi(t,x)\psi(t,x),\nn
\ea
such that $S_<$ depends only on $\psi(t_{in}<t'\leq t)$ and $\hat\psi(t_{in}\leq t'<t)$ while $S_>$ involves $\psi(t<t'\leq t_f)$ and $\hat\psi(t\leq t'<t_f)$. For any given time $t$ we can specify the state of the system by the Grassmann element
\be\label{G4}
g(t)=g\big[\psi(t)\big]=\int{\cal D}\psi(t'<t){\cal D}\hat\psi(t'<t)e^{-S_<}g_{in}.
\ee
It belongs to the Grassmann algebra which can be constructed from $\psi_\gamma(t,x)$ for a fixed $t$. The conjugate element obeys
\be\label{G5}
\hat g(t)=\hat g\big[\hat\psi(t)\big]=\int{\cal D}\psi(t'>t){\cal D}\hat\psi(t'>t)\hat g_f
e^{-S_>}.
\ee
It obtains from $g(t)$ by the replacement $\psi(t)\to\hat\psi(t)$ and total reordering. The proof of this property requires the hermiticity of $S_M$ \cite{CWF}. The quantity $g(t)$ plays the role of the wave function for a quantum state in the Grassmann formulation.

As in the main text, the Grassmann element $g(t)$ can be expanded in a complete basis of the Grassmann algebra which is constructed from Grassmann variables $\psi_\gamma (t,x)$ at a fixed time $t$,
\be\label{G6}
g(t)=q_\tau(t)g_\tau.
\ee
For $L^3$ space points we have $B=4L^3$ Grassmann variables and therefore $2^B$ independent basis elements $g_\tau$. Each $\tau$ will be associated with a state of a classical statistical ensemble, which is characterized by an ordered chain of $B$ bits taking the values $1$ (occupied or spin up) or $0$ (empty or spin down), $\tau=[n_\alpha]=\big [n_\gamma(x)\big]$. For every $\psi_\alpha$ appearing in $g_\tau$ we take the bit $\alpha$ empty, while the bit $\alpha$ is occupied if $\psi_\alpha$ does not appear in $g_\tau$. As an example, we can associate the Grassmann element $g_\tau=\psi_1\psi_4\psi_5\psi_7\dots$ with the bit chain $\tau=[0,1,1,0,0,1,0\dots]$. In our case, the states $\tau$ of the classical statistical  ensemble can be identified with a three dimensional array of four ``species'' of bits in a computer, or with the possible states of a four-component Ising model in three dimensions. In terms of the ``classical wave function'' $\{q_\tau(t)\}$ the probability distribution $\big\{p_\tau(t)\big\}$ of the classical statistical ensemble is given by
\be\label{G7}
p_\tau(t)=q^2_\tau(t).
\ee
The positivity of $p_\tau$ is obvious and the normalization
\be\label{G7A}
\sum_\tau p_\tau(t)= \sum_\tau q_\tau(t)q_\tau(t)=
{\cal N}_p(t)=1
\ee
can be written in the formulation of this appendix as 
\ba\label{G8}
{\cal N}_p(t)&=&\int{D}\psi(t)\hat g(t)g(t),\\
\int {D\psi}(t)&=&\int {\cal D}\psi(t){\cal D}\hat\psi(t)\exp \big\{\int_x\hat\psi_\gamma(t,x)\psi_\gamma(t,x)\big\}.\nn
\ea
Here we use the definition of the conjugate Grassmann element
\be\label{G9}
\hat g(t)=q_\tau(t)\hat g_\tau.
\ee
In eq. \eqref{G9} the Grassman element $\hat g_\tau$ is conjugate to the basis element $g_\tau$, obeying 
\be\label{G10}
\int D\psi\hat g_\tau g_\rho=\delta_{\tau\rho}.
\ee
(For the example $g_\tau=\psi_1\psi_4\psi_5\psi_7\dots$ one has $\hat g_\tau=\dots \hat\psi_7\hat\psi_5\hat\psi_4\hat\psi_1.$) If $\hat g(t)$ as defined by eq. \eqref{G5} coincides with the conjugate of $g(t)$ as defined by eq. \eqref{G10}, it is easy to see that ${\cal N}_p(t)$ is independent of $t$. The definitions \eqref{G4} and \eqref{G5} imply ${\cal N}_p(t)=Z$. In turn, it is sufficient to choose a properly normalized $g_{in}$ with ${\cal N}_p(t_{in})=1$ in order to guarantee $Z=1$. 

\bigskip\noindent
{\bf   3. Time evolution}

\medskip
We have seen in the main text that for the action \eqref{N1}-\eqref{C.8} the time evolution of the real vector $\{q_\tau(t)\}$ is given by a rotation, which guarantees the normalization \eqref{G7A} if $g_{in}$ is properly normalized. It is instructive to show the ``unitary time evolution'' also in the formalism of this appendix in terms of conjugate Grassmann variables $\hat\psi$. We start with 
\ba\label{30A}
g(t+\epsilon)&=&\int d\psi (t) d\hat\psi(t)\exp \{-L(t)\}g(t)\\
&=&\int d\psi(t)\delta\big (\psi(t)-\psi(t+\epsilon)\big)e^{\epsilon{\cal K}}g(t),\nn
\ea
with ${\cal K}$ given by eq. \eqref{P37}. Here the Grassmann integration extends over all variables $\psi_\gamma(t,x)$ and $\hat\psi_\gamma(t,x)$ at a given time $t$. Similarly, the Grassmann $\delta$-function extends over all $\psi_\gamma(t,x)$. We recall that $g(t)$ and $\exp\{\epsilon{\cal K}\}g(t)$ depends on the variables $\psi(t)$, while $g(t+\epsilon)$ involves $\psi(t+\epsilon)$. The shift in variables is provided by the $\delta$-function, which corresponds to the first term in $L(t)$ \eqref{G1} (not involving $T_k\partial_k$). The infinitesimal formulation with continuous time avoids the ordering problems of the operator $\exp \{\epsilon{\cal K}\}$ by neglecting corrections $\sim \epsilon$. This holds provided that the wave function $q_\tau(t)$ depends on time in a sufficiently smooth way. (For details of the construction cf. ref. \cite{CWF}.)

For our case we infer from eq. \eqref{G1} 
\ba\label{30B}
g(t+\epsilon)&=&\prod_{x,y}\Big[\int d \psi_\gamma(t,x)\int d\hat\psi_\gamma(t,x)\\
&\times&\exp \big \{\hat\psi_\gamma(t,x)\big[\psi_\gamma(t,x)-B_\gamma\big]\big\}\Big]g(t),\nn
\ea
with $B_\gamma$ depending linearly on $\psi(t+\epsilon)$
\be\label{30C}
B_\gamma(t+\epsilon,x)=\tilde A\psi_\gamma(t+\epsilon,x)-\epsilon(T_k)_{\gamma\delta}\partial_k\psi_\delta(t+\epsilon, x).
\ee
The integral over $\hat\psi_\gamma$ yields a $\delta$-function, such that
\ba\label{30D}
g(t+\epsilon)&=&\prod_{x,y}\big[\int d\psi_\gamma(t,x)\delta\big(\psi_\gamma(t,x)-B_\gamma(t+\epsilon,x)\big)\nn\\
&\times&g\big[\psi_\gamma(t,x)\big]=g[B_\gamma].
\ea
In other words, $g(t+\epsilon)$ obtains from $g(t)$ by replacing each Grassmann variable $\psi_\gamma(t,x)\to B_\gamma(t+\epsilon,x)$. As in eq. \eqref{C.9} $B_\gamma$ is a rotation of $\psi_\gamma$, such that $g_\tau(t+\epsilon)$ obtains from $g_\tau(t)$ by a rotation, and similarly for $q_\tau(t+\epsilon)$. In the formalism with conjugate spinors the time evolution of the expectation value is given by
\be\label{G20}
\partial_t\kl A\kr=\kl q[\hat A,K]q\kr=\int D\psi\hat g[\cA,\cK]g,
\ee
with
\be\label{37A}
\partial_t\hat g=-\hat\cK^T\hat g,
\ee
and $\hat \cK^T$ obeying
\be\label{G21}
\int D\psi\hat \cK^T\hat gf=\int D\psi\hat g\cK f
\ee
for arbitrary Grassmann elements $f$. 

\bigskip\noindent
{\bf  4. Symmetries}

\medskip
The formalism with conjugate spinors is convenient for the discussion of those symmetries where Grassmann variables at even and odd time points transform differently. We will concentrate on continuous symmetries of the action which involve transformations between $\psi$ and $\hat\psi$. Grouping the four-component vectors $\psi_\gamma$ and $\hat\psi_\gamma$ into an eight component real vector $\tilde \psi$, 
\be\label{71E}
\tilde\psi={\psi \choose\hat\psi},
\ee
the action is invariant under infinitesimal transformations
\be\label{71F}
\delta\tilde\psi=\epsilon_VV\tilde\psi
\ee
if the $8\times 8$ matrix $V$ obeys (with $T_k=diag(T_k,T_k))$
\be\label{71G}
[V,T_k]=0~,~V^T=-CVC~,~C={0,1\choose 1,0}.
\ee
Independent matrices obeying the condition \eqref{71G} are
\be\label{71GA}
V=\{\tilde I~,~C\tilde I~,~N~,~\tilde IW\},
\ee
where $\tilde I=diag(\tilde I,\tilde I)$, $\tilde I C=C\tilde I$, and
\be\label{71H}
N=\left(\begin{array}{rr}
-1,&0\\0,&1
\end{array}\right)~,~
W=\left(\begin{array}{rr}
0,&-1\\1,&0
\end{array}\right).
\ee
The transformations generated by $V$ \eqref{71GA} commute with the Lorentz transformations. For $V=\tilde I$ we recover eq. \eqref{71B}, while the antisymmetric matrix $V=C\tilde I=-(C\tilde I)^T$ describes an abelian $SO(2)$ rotation between $\psi$ and $\hat\psi$. 

In analogy to eq. \eqref{71E} we can group the one-particle and one-hole wave functions $q$ and $\hat q$, or $\varphi$ and $\hat\varphi$, into a common wave function $\tilde \Phi$. For the complex four component spinor $\tilde \Phi$
\be\label{71I}
\tilde\Phi={\varphi\choose\hat\varphi}~,~\hat\varphi=
{\hat q_1+i\hat q_2\choose \hat q_3+i\hat q_4}
\ee
the first two symmetries are given by
\be\label{71J}
\tilde \Phi'=e^{i\alpha}\tilde\Phi~,~\tilde\Phi'=e^{i\beta C}\tilde\Phi.
\ee
The other two possibilities for $V$ are symmetric, $N^T=N~,~(\tilde I W)^T=\tilde I W$, and therefore do not belong to the generators of eight-dimensional rotations. For example, the transformation associated to the generator $N$ describes a scaling
\be\label{71K}
\psi\to e^{-\gamma}\psi~,~\hat\psi\to e^\gamma\hat\psi.
\ee

The action of $N$ and $\tilde I$ does not mix $\psi$ and $\hat\psi$. It can therefore be defined for the Grassmann basis elements $g_\tau$ and by extension for arbitrary Grassmann elements $g$. For a Grassmann element $g_\tau=\prod\limits^{\hat n}_{k=1}\psi_{\gamma_k}(x_k)$ the infinitesimal transformation associated to $N$ reads
\be\label{71L}
\delta g_\tau=-\hat n\epsilon_N g_\tau=-(B-n)\epsilon_N g_\tau,
\ee
such that $Ng_\tau=(n-B)g_\tau$. We can identify $N$ with the (shifted) particle number operator \eqref{H5}, $N={\cal N}-B$. The scaling symmetry \eqref{71K} is associated to the conserved particle number. 
\ba\label{B17}
&&\hat\psi(x)\to \bar{\cal P}\hat\psi(x)\nn\\
&&\big (\bar{\cal P}\hat\psi(x)\big)_\gamma=(\gamma^0)_{\gamma\delta}
\hat\psi_\delta(-x),
\ea
The action \eqref{F1} is invariant under a particle-hole conjugation transformation which exchanges $\psi_\gamma(x)$ and $\hat\psi_\gamma(x)$,
\be\label{K1}
C_{ph}:\quad \psi_\gamma(x)\leftrightarrow\hat\psi_\gamma(x).
\ee
This symmetry commutes with the Lorentz transformations. It reflects the invariance of the evolution equation for the probability distribution of the classical  statistical ensemble under the exchange of empty and occupied bits.

\end{document}